\documentclass[twocolumn,showpacs,preprintnumbers,amsmath,amssymb]{revtex4}
\usepackage{psfig,epsf,graphics,graphicx,bm}
\begin{document}

\title{ARPES spectra and the single particle spectral weight \\
 of cuprates in the bond-ordered, bond-centered stripe phase}

\author{P. Wr\'obel$^{1,2}$, A. Maci\c{a}g$^{2}$ and  R. Eder$^3$}
\affiliation{$^1$ Max Planck Institute for the Physics of Complex Systems,
 D-01187 Dresden, Germany }
\affiliation{$^2$ Institute for Low Temperature and Structure
Research, P. 0. Box 1410, 50-950 Wroc{\l}aw 2, Poland}
\affiliation{$^3$ Forschungszentrum Karlsruhe, IFP, P.O. Box 3640, D-76021
Karlsruhe, Germany}

\begin{abstract}
The electronic structure and the  single-particle spectral density of a stripe
 array formed by ladder-like domain walls (DWs) and by antiferromagnetic (AF)
 domains of width 2 lattice spacings are computed and compared with ARPES
 spectra from some doped cuprates belonging to the 214 family of compounds.
We assume that bond order is formed on legs in DWs and that the phase
 of the sublattice magnetization changes by $\pi$ across each DW.
The intensity map plotted in the coordination frame momentum-energy
 reproduces quite well the ARPES spectra obtained at the doping level
 of $15\%$. 
We consider this agreement as an argument for a scenario of coexisting
 bond-ordered regions and AF regions in the stripe phase of moderately
 doped cuprates.
\end{abstract}
\pacs{71.10.Fd, 74.20.Mn, 74.72.-h, 79.60.-i}

\maketitle
\section{introduction}
A tendency towards spin and charge ordering in cuprates has been observed in 
 results of several neutron scattering experiments.
Low frequency spin fluctuations observed by many groups at incommensurate 
 wave vectors as relatively sharp peeks in the magnetic structure factor
  \cite{thurston89,tranquada95,yamada95}
 have been interpreted as being due to stripe fluctuations
 \cite{kivelson03}.
Anisotropy observed in resistivity \cite{ando02} also points at stripe
 formation.
Stripe order appears to have an impact on phonon mediated heat transport
 \cite{hess99}.
Measurements performed by means of NMR and NQR techniques demonstrate
 the emergence of slow spin fluctuations whose appearance is correlated with 
 pinning  of charge-modulations \cite{hunt99}. 
The distribution of nearest neighbor bond lengths deduced from neutron powder
 diffraction data \cite{bozin00,braden01} and measured by means
 of extended  x-ray-absorption fine  structure spectroscopy \cite{bianconi96}
 agrees with expectations based on a scenario of
 lattice response to local charge-stripe order.
An indication of stripe order may by also found by analyzing the shape of ARPES
 spectra and their evolution with doping. 
For example, it is natural to expect the Fermi surface to be flat in
 the stripe phase in the anti-nodal region, near the vectors
 $(\pm\pi,0)$ and $(0,\pm\pi)$ \cite{kivelson03}.
In addition, quasi-one-dimensionality of the system should bring about
 the depletion of spectral weight in the nodal regions,
 near the Brillouin zone diagonals.
Much research has been done to check if these hypotheses are true
 \cite{zhou99,zhou01,fedorov99,campuzano03,kivelson03,damascelli03}.
It seems that some general structure of spectra may be definitely attributed
 to an underlying quasi-one dimensional electronic structure
 \cite{carlson01,carlson03,ichioka99,zacher00,zacher02,eroles01,granath01,granath02,fleck00,web,markiewicz00}.

In this paper we will concentrate on doped $La_{2-x} Sr_x Cu O_4$ (LSCO)
 compounds, slightly above the doping level $1/8$ at which, it is believed,
 a tendency towards nanoscale phase separation seems to be evident in the
 the ARPES results \cite{damascelli03,ino00,zhou99}.
In ARPES intensity maps obtained by integrating over 30 $meV$ 
 at the Fermi energy the spectra from 15\% doped 
 $La_{2-x-y}Nd_y Sr_x Cu O_4$ (Nd-LSCO) \cite{zhou01}, patches
 with high intensity may be seen in anti-nodal regions.
Appreciable spectral weight is also detected at the Fermi energy 
 in nodal regions.
A similar pattern of the spectral density at the Fermi energy was
 predicted by  a phenomenological theory of disordered charge stripes and 
 anti-phase spin domains \cite{salkola}.
Unfortunately, this theory does not discuss the origin of renormalized 
 hopping terms in the effective one-body Hamiltonian and does not explain
 the relation of this renormalization with the underlying magnetic structure.
The evolution of the spectral weight as a function of doping  has been
 analyzed for  the stripe phase by means of the cluster perturbation technique
 (CPT) in the framework of the microscopic $t$-$J$ and Hubbard models
 \cite{zacher02}.
This theory captures quite well the general trend of this development.
Nevertheless, it seems that the spectral weight maps derived within this
 approach for the doping level at and above 12.5\% do not show continuous well
 developed high-intensity straight patches bridging anti-nodal regions.
Such structures are experimentally seen in nodal regions \cite{zhou01}.
In addition, that theory does not provide us with much information about
 the magnetic structure of stripes.
The remarks made above seem to suggest that some understanding of the relation
 between the single particle spectral weight of Nd-LSCO at the filling level
 about $1/8$ and the formation of the stripe phase is missing.
A phase which we may expect to emerge in a natural way in weakly doped 
 AFs is a bond-ordered state \cite{anderson}.
Recently, an exact diagonalization of the $t$-$J$ model ($tJ$M) at a finite
 cluster has been performed to study stripe formation.
It has been shown that the cluster geometry change, from the standard tilted
 square form of the 20 sites cluster to the rectangular one, $5\times4$
 cluster induces the formation of a ground state with pronounced
 stripe-like charge inhomogeneities \cite{ederohta04a}.
The distribution of peaks in the single-particle spectral weight, which has 
 been calculated by means of the same method \cite{eder04b}, resembles 
 experimental  ARPES spectra from $La_{1.28}Nd_{0.6}Sr_{0.12}CuO_4$ 
 \cite{zhou99}.
In particular, the theory captures quite well the strong dispersion along
 the $(0,0)$ - $(\pi,0)$ line.
The spreading of the quasiparticle-peaks in the single-particle spectrum
 obtained by means of the exact diagonalization is in good agreement with
 results of an additional calculation performed by means of a different method
 which is the bond operator theory for the spin-Peierls phase.
Unfortunately, neither the numerical approach nor the analytical method
 reproduce the flattening of the experimental band near the point $(\pi,0)$ and 
 both of them fail to
 explain the emergence of the spectral weight at the nodal region, in the form
 of a straight patch which is observed at the Fermi energy in the experiments. 
Another analysis of the spectral weight in doped AFs, based on the exact
 diagonalization of a small cluster concerns the $tJ$M with inhomogeneous
 terms locally breaking the translational invariance and the spin-rotational
 SU(2) symmetry \cite{eroles01}.
The pattern of the spectral weight at the Fermi energy obtained by means
 of this method resembles ARPES spectra from Nd-LSCO at the doping level
 $12\%$ and does not show enhanced intensity at nodal regions.
This enhancement observed in experiments is the
 manifestation of remnant two-dimensionality (2D) in the stripe system.
Thus, also this theory seems not account for the shape of ARPES spectra from
 Nd-LSCO or LSCO at the doping level slightly higher than $1/8$.

Results of inelastic neutron scattering (INS) experiments indicate that 
 pronounced AF correlations exist in the stripe phase \cite{tranquada95}.
On the other hand, the relevance of bond-order has been recently demonstrated
 by means of the same experimental method  \cite{hayden04}.
In a recent work  \cite{wrob04} we have suggested that the coexistence 
 of bond-order with long range AF order may take place in the stripe phase.
The scenario of nanoscale phase separation is realized
 by means of this coexistence.
We have shown in the framework of this concept that above the doping
 level $1/8$, when the distance between stripe axes is 4 lattice spacing a 
 bond-centered stripe with spin-Peierls order on these bonds is more stable
 than a site-centered stripe.
In addition it is also known about the site-centered stripe that it is
 more stable than a homogeneous system of holes created in the
 homogeneous AF \cite{chernyshev00,web}.
The mechanism of coexistence between AF order and spin-Peierls order
 in the stripe phase is based on lowering the kinetic energy of holes
 moving without confinement in hole-rich stripes which are formed in the 
 form of two-leg ladder-like domain walls (DWs) between AF hole-poor 
 domains in which the exchange energy decreases \cite{wrob04}.

In the next section, in the framework of a scenario for nanoscale phase 
 separation and coexistence of AF long range order with bond-order, we will
 derive an effective tight-binding Hamiltonian describing a quasiparticle
 propagating in such a spin background.
Next we will calculate the part of the spectral function which is accessible
 to measurements in photoemission experiments. 
Finally we will discuss results of the calculation.
They seem to show characteristic features that may be seen in ARPES spectra
 of Nd-LSCO slightly above the doping level $1/8$.

\section{Band structure of a stripe system with coexisting bond and AF orders}
\begin{figure}
 \unitlength1cm
\begin{picture}(6.5,7)
\epsfxsize=8.0cm
\put(-0.6,0){\epsfbox{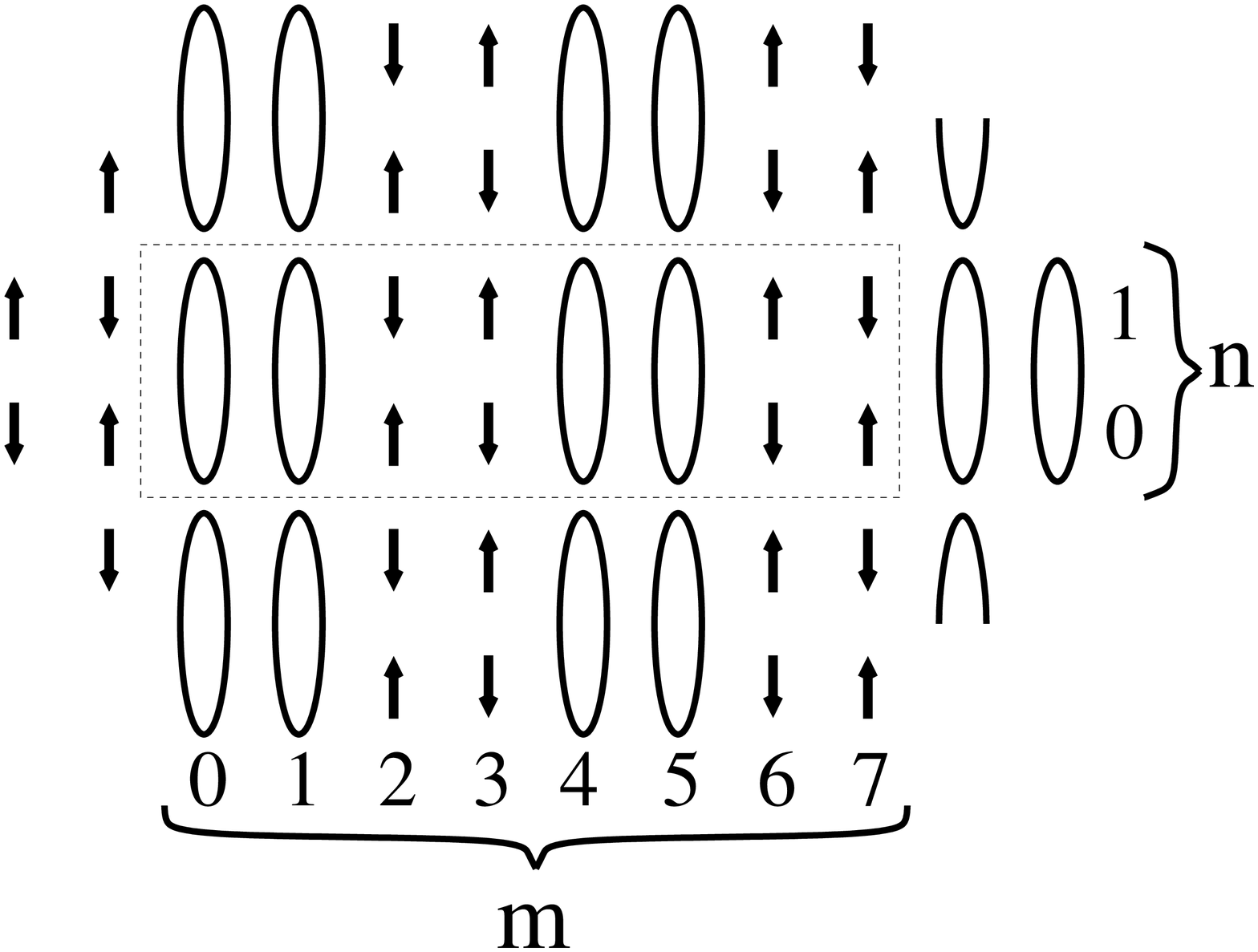}}
\end{picture}
\caption{Elementary cell of the underlying spin structure assumed in the
 calculation (inside the dashed rectangle). Ovals represent singlets.}
\label{fig1}
\end{figure}

\begin{figure}
 \unitlength1cm
\begin{picture}(6.5,7)
\epsfxsize=8.0cm
\put(-0.6,0){\epsfbox{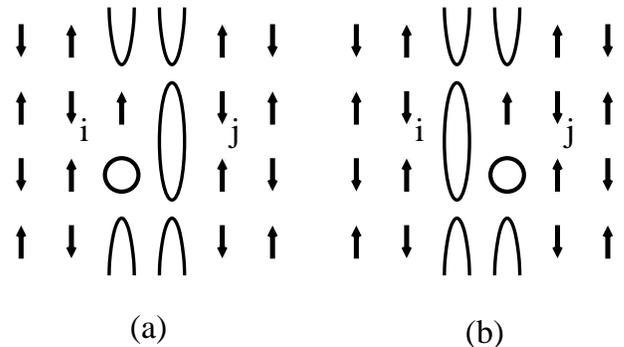}}
\end{picture}
\caption{The mechanism which gives rise to FM coupling between nearest spins
 which belong to different domains. Notice, that these spins would belong
 to the same AF sublattice if the system were homogeneously ordered.}
\label{fig2}
\end{figure}

\begin{figure}
 \unitlength1cm
\begin{picture}(6.5,4)
\epsfxsize=9.0cm
\put(-1.3,0){\epsfbox{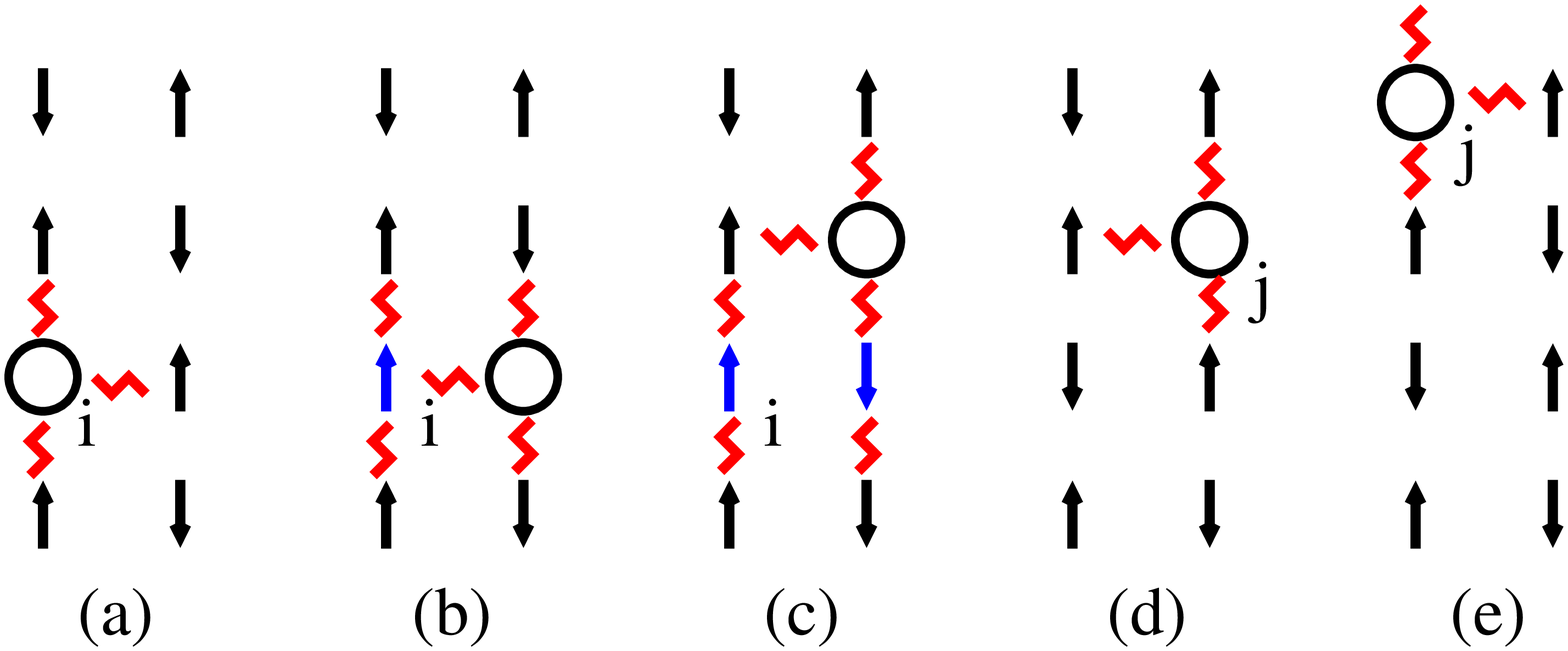}}
\end{picture}
\caption{Graphical representation of some states involved in the process of
 hole propagation inside an AF domain.
 Zig-zag lines represent ``broken bonds'' contribution from which to
 Ising energy is higher by $J/2$ than contribution from bonds
 occupied by two antiparallel spins.}
\label{fig3}
\end{figure}

\begin{figure}
 \unitlength1cm
\begin{picture}(6.5,15)
\epsfxsize=7.3cm
\put(-0.2,0){\epsfbox{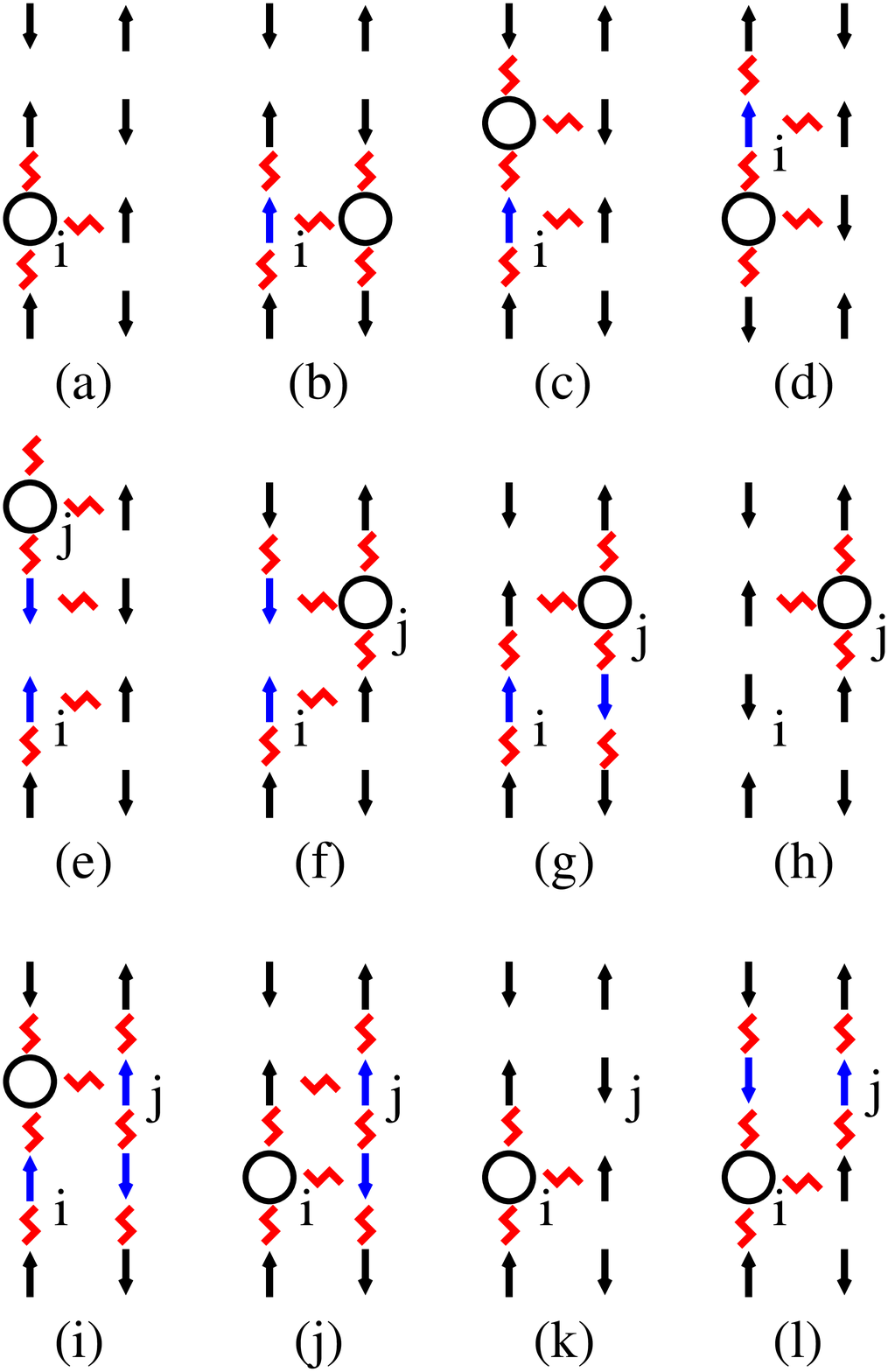}}
\end{picture}
\caption{Illustration of processes contributing to some terms
 in the effective Hamiltonian.
 These terms define the on-site energy and the quasiparticle hopping
 between sites which belong to AF domains.}
\label{fig4}
\end{figure}

The $tJ$M in the framework  of which we perform the calculation is
\begin{equation}
H=-\sum_{i,j} t_{ij} c_{i,\sigma}^\dagger c_{j,\sigma}
+J \sum_{\langle i,j \rangle}(\mathbf{S}_i \mathbf{S}_j-\frac{n_i n_j}{4}).
\label{hamorg}
\end{equation}
The states in which any site is doubly occupied have been excluded
 by definition from the Hilbert space in which that model acts.
$\langle i,j \rangle$ represents pairs of nearest neighbor (NN) sites.
$\vec{S_i}$ and $n_i$ denote the operators of electron spin and density
 at site $i$ respectively.
$t$ represents the hopping matrix element $t_{i,j}$ between NN sites in
 the square lattice on which  the $tJ$M is defined, $t'$- the hopping matrix
 element $t_{i,j}$ between second NN sites, and $t''$ - the hopping matrix
 element between third NN sites.
The rest of $t_{i,j}$ is zero.
We concentrate on the doping level about $1/8$ at which the distance between
 axes of nearest  stripes is 4 lattice spacings and AF correlations
 seem to be of long range \cite{tranquada95}.
Our previous analysis has provided convincing arguments that in such a case
 the stripe takes the form of a two-leg ladder-like DW which separates 
hole-poor
 AF domains \cite{wrob04}.
The phase of sublattice magnetization in domains changes 
 by $\pi$ across a DW.
Each DW is in the spin-Peierls state with singlets formed on legs. 
The underlying spin structure of the stripe system at the doping level 
 about $1/8$ has been presented in Fig.\ref{fig1}.
A natural question arises, what is mechanism which gives rise to long range
 AF correlations, if AF domains are separated by DWs which consist of singlets?
The structure depicted in Fig.\ref{fig1} emerges in the presence 
 of holes only.
That presence is induced by doping.
The creation of a hole gives rise to appearance of an uncompensated spin in 
 the DW. 
This spin may be parallel or antiparallel to a nearest spin in the neighbor
 domain.
The weight of states in which these spins are antiparallel, Fig.\ref{fig2}(a),
 (b), is higher, 
 because such a configuration is preferred by AF coupling between NN sites.
States depicted by Fig.\ref{fig2}(a) and (b) are coupled by the hopping term
 in the Hamiltonian.
Since their weight is higher than the weight of states in which an
 uncompensated spin in the DW and the nearest spin in the domain are parallel,
 the hopping term which transforms the state depicted by Fig.\ref{fig2} (a)
 into the state depicted by Fig.\ref{fig2}(b), mediates effective 
 ferromagnetic (FM) coupling between sites $i$ and $j$. 
Some quantum fluctuations will be present in the underlying spin 
 background of the stripe system. 
They may take the form of triplet excitations on bonds in DWs and multimagnon
 excitations in domains.
Fig.\ref{fig4} in the previous work \cite{wrob04} contains some examples
 of quantum fluctuations in the spin background.
They contribute a lot to the energy of the system.
On the other hand, it seems that quantum fluctuations merely renormalize
 the shape of quasiparticle dispersion for a given spin background and
 do not change the picture qualitatively.
This has been demonstrated in the case of a hole propagating in the AF spin
 background \cite{eder91},
 as well as in the case of bond ordered  two-leg ladders \cite{eder98,brenig01}
 and bond-ordered 2D systems \cite{eder04b}. 
When it is necessary we will take into account the 
 influence of quantum fluctuations on the distribution of spectral weight.
A scenario which underlies the calculation which we are going to outline next,
 is based on the assumption that hole motion inside ladder-like 
 DWs is governed by the exchange of positions between a hole-fermion pair on
 a bond and a singlet on a nearby bond \cite{eder98}.
This exchange is mediated by the hopping terms in the initial 
 Hamiltonian (\ref{hamorg}).
We also assume that a hole propagates in the AF spin background as a spin
 polaron \cite{chernyshevwood02}.
In the calculation we take into account a simplest form of 
 coupling which moves a hole between a bond ordered DW and an AF domain. 
This form of coupling originates in the hopping term of the initial 
 Hamiltonian.
Before we proceed to construct an effective Hamiltonian which describes
 the motion of a quasiparticle in the underlying spin background,
 we will present some formulas which will be useful for that purpose.
A singlet on two sites  ``$L$'' and ``$U$'', is created in the 
 vacuum by the operator
\begin{equation}
s^\dagger_{LU}=\frac{i}{\sqrt{2}} [\sigma^0 \sigma^y]_{\alpha \beta} c^\dagger_{L \alpha} c^\dagger_{U \beta}.
\label{sin}
\end{equation}
$\sigma^0$ is the two-dimensional identity matrix and $\sigma^a$, $a=x,y,z$,
 are Pauli matrices.
The summation over repeating Greek indices is assumed.
Three operators which are components of the vector 
 $\mathbf{t}^\dagger_{LU}$  create three triplet states on sites $L$ and $U$,
\begin{equation}
\mathbf{t}^\dagger_{LU}=\frac{i}{\sqrt{2}} [{\bm{\sigma}}\sigma^y]_{\alpha \beta} c^\dagger_{L \alpha} c^\dagger_{U \beta}.
\label{trip}
\end{equation}
A formula which we will often use later is
\begin{eqnarray}
-c_{U'\sigma}^\dagger c_{U\sigma}(s^\dagger_{LU} c^\dagger_{L' \gamma}) |0\rangle
&=&(\frac{1}{2} s^\dagger_{L'U'} c^\dagger_{L \gamma}
\nonumber \\
&+& \frac{1}{2} \mathbf{t}^\dagger_{L'U'} {\bm{\sigma}}_{\alpha \gamma} c^\dagger_{L \alpha})|0\rangle.
\label{hsex}
\end{eqnarray}
The left side of (\ref{hsex}) together with the right-side term which 
 contains a singlet at the pair of sites $L'$ and $U'$ represent the exchange
 of a singlet and a hole-fermion pair between two bonds, which is mediated by
 the hopping term in the initial Hamiltonian.
As we have already stated before, we will neglect the creation of triplets 
 on bonds, however this process also takes place during hole hopping.
The creation of a triplet is represented by the second term  at the right side 
 of (\ref{hsex}).
We also need formulas which represent the hopping of a hole from
 a single site $i$ which belongs to an AF domain to a site which belongs to
 a bond occupied by a singlet and vice versa
\begin{eqnarray}
-c^\dagger_{i \sigma} c_{U \sigma} (s^\dagger_{LU})|\Omega\rangle &=&
-\frac{i}{\sqrt{2}} \sigma^y_{\alpha \beta} c^\dagger_{L \alpha} c^\dagger_{i \beta}|0\rangle,
\label{defs1}
\\
-c^\dagger_{U \sigma} c_{i \sigma} (c^\dagger_{L \alpha} c^\dagger_{i \beta}) 
|0\rangle &=& ( -\frac{i}{\sqrt{2}}\sigma^y_{\alpha \beta} s^\dagger_{LU} 
\nonumber \\
&+& \frac{i}{\sqrt{2}}[\sigma^y {\bm{\sigma}}]_{\alpha \beta} \mathbf{t}^\dagger_{LU}) |0\rangle.
\label{defs2}
\end{eqnarray}
When we start to analyze contributions from quantum fluctuations in
 the ground state of the system to the spectral function we will need
 followings formulas
\begin{eqnarray}
\mathbf{S}_L \mathbf{S}_i (s^\dagger_{LU}c^\dagger_{i \beta}|\phi\rangle
&=&\frac{1}{2} \bm{\sigma}_{\alpha \beta} 
(\mathbf{t}^\dagger_{LU}c^\dagger_{i \alpha})|0\rangle,
\label{act1}
\\
\mathbf{S}_U \mathbf{S}_i (s^\dagger_{LU}c^\dagger_{i \beta}|\phi\rangle
&=&-\frac{1}{2} \bm{\sigma}_{\alpha \beta} 
(\mathbf{t}^\dagger_{LU}c^\dagger_{i \alpha})|0\rangle.
\label{act2}
\end{eqnarray}
We begin the derivation of the effective Hamiltonian describing hole 
 propagation  in the spin background depicted by Fig.\ref{fig1} with 
 outlining the mechanism of hole propagation inside an AF domain.
A hole created and moving in the N\'eel background shifts spins between 
 different sublattices and creates defects in the AF structure,
 Fig.\ref{fig3} (a)-(c).
Such a process gives rise to an increase of the Ising energy. 
This rise is roughly speaking proportional to the length of a path along which
 the hole has traveled, which means that a tendency towards hole confinement 
 appears \cite{kane88}.
In order to take into account such a tendency we will analyze hole
  motion in the framework of a basis which consists of states representing 
 holes confined in the AF background by linear defects (strings)
 left behind by moving holes on their way.
We call these states spin polarons.
A wave function representing a confined spin polaron at a site $i$
 in an AF domain is a 
 combination of states which are created from the state
 $c_{i,\uparrow (\downarrow)}|N\rangle$ by the NN hopping term when a hole 
 created at the site $i$ starts to move,
\begin{equation}
|\Psi_i\rangle=\sum_{{\cal P}_i} \alpha _{l({\cal P}_i)}|{\cal P}_i\rangle,
\label{polar}
\end{equation}
$|N\rangle$ is the N\'eel state in the domain, $|{\cal P}_i\rangle$ 
 denotes a state obtained by hopping along a path ${\cal P}_i$ of a hole 
 created at the site $i$.
$\alpha_{l({\cal P}_i)}$ is the amplitude of this state.
We have assumed for simplicity that $\alpha_{l({\cal P}_i)}$
 depends solely on the length $l({\cal P}_i)$ of the  path ${\cal P}_i$.
The length of a path or of a string state is defined as the number of hops
 needed to form a given string state from a state representing a hole created
 in the AF spin medium.
At the first stage of the analysis we take into account only the hopping
 between NN sites because $t\gg J,t',t''$.
Processes related to hopping between further neighbors and processes related 
 to swapping antiparallel spins by the transversal term in the exchange 
 interaction will be considered later as a perturbation.
A hole moving inside a domain may make its first step in $(z-1)$ directions.
$z$=4 is the coordination number of the square lattice.
There are in principle $(z-2)$ direction choices of each next hop, if the hole
 moves without retracing inside the domain.
On the square lattice there are $z$ choices for the direction of the first
 step and $(z-1)$ for the direction of further hops during the non-retractable
 motion.
These number gets reduced by one for the domain formed by two chains of sites.
Thus if we neglect some details, as for example path crossing, we may write,
\begin{equation}
\langle \Psi_i|\Psi_i\rangle=\alpha_0^2+(z-1)\sum_{\mu = 1}(z-2)^{\mu -1} \alpha^2_\mu.
\label{norm}
\end{equation}
Each prefactor in (\ref{norm}) at the square $\alpha^2_\mu$ 
 represents the number of different paths with the length $\mu$.
We calculate the energy of the spin polaron state $|\Psi_i\rangle$, $\varepsilon_1$,
 which is given by the expectation value of a trial Hamiltonian $H_0$, 
 $\langle \Psi_i|H_0|\Psi_i\rangle$, with the assumption, that the motion of 
 a hole which has started from the site $i$, is restricted to the interior
 of the domain, and that the contribution from the interaction term is
 restricted to the Ising part 
 $\sum_{\langle i,j \rangle} (S_i^z S_j^z - \frac{n_i n_j}{4})$,
\begin{eqnarray}
\langle \Psi_i|H_0|\Psi_i\rangle &=& [3\alpha_0^2+(z-1)\sum_{\mu=1}(z-\mu)^{\mu -1}
(4+\mu)\alpha^2_\mu]\frac{J}{2}
\nonumber\\
&+&2(z-1)\sum_{\mu=0}(z-2)^\mu \alpha_\mu \alpha_{\mu+1}t.
\label{pH0p}
\end{eqnarray}
The first term in (\ref{pH0p}) basically counts the number of ``broken bonds'',
 that are not occupied by a pair of antiparallel spins, in which case the Ising
 contribution to the energy of that bond is higher by $J/2$ than in the case 
 if it were occupied by a pair of antiparallel spins.
The second term in (\ref{pH0p}) is the contribution from the hopping operator
 to the spin polaron energy. 
The prefactors appearing in this term represent the number of paths with
 a given length multiplied by the number of directions in which these paths
 may be extended.
The appearance of the factor 2 is related to the fact that the hopping  which
 couples paths of length $\mu$ and $\mu+1$ may take place forth and back.
The values of these paths $\alpha_\mu$ can be found by minimizing 
 $\langle \Psi_i |H|\Psi_i \rangle$ under the constraint
 $\langle \Psi_i |H|\Psi_i \rangle=1.$
After we have constructed spin polarons which are formed in AF domains we
 are able to present the full basis of single-particle states.
The underlying spin background which has been presented in Fig.\ref{fig1} plays
 the role of the vacuum $|\Omega \rangle$ for hole-like quasiparticles which
 propagate  in this background.
$|\Omega \rangle$ has been obtained by acting on the absolute vacuum
 $|0\rangle$, in which no particles are present, with
 a product of operators like $s^\dagger_{LU}$ creating singlets on bonds
 connecting sites $L$ and $U$ and operators $c^\dagger_{i \sigma}$ creating
 spins in domains according to the pattern shown in Fig.\ref{fig1}.
New fermionic positively charged  operators $h^\dagger_{i \sigma}$
 create single particle hole-like states from the vacuum for holes
 $|\Omega\rangle$.
The action of the operator $h^\dagger_{L \sigma}(h^\dagger_{U \sigma})$ on
 $|\Omega\rangle$, where the site $L(U)$ belongs to a ladder-like DW,
 exchanges the operator $s^\dagger_{LU}$ in the 
 product defining $|\Omega\rangle$
 by the operator $c^\dagger_{U \sigma} (c^\dagger_{L \sigma})$, which means
 that instead of a singlet on the bond connecting sites $L$ and $U$ there is 
 a hole on the site $L(U)$ and spin $\sigma$ on the site $U(L)$.
The action of the operator $h^\dagger_{i \sigma}$ on a site $i$ which
 belongs to an AF domain, creates a spin polaron $|\Psi_i\rangle$
 in that domain.
The spin polaron is a combination of some states, the amplitudes of which are
 given by prefactors $\alpha_\mu$.
These states include $c_{i \bar{\sigma}}|\Omega\rangle$ and states obtained
 by applying consecutively the NN inside-domain hopping term to the state
 $c_{i \bar{\sigma}}|\Omega\rangle$.
During this process, the hole hops without retraces and $\mu$ 
 is $l({\cal P}_i)$, the length of the path ${\cal P}_i$, along which the hole
 has traveled to form a given component state of 
 $|\Psi_i\rangle$.
A label which we will use to mark the fermionic operator creating either 
 a bond hole or a hole-like spin polaron at a give site is 
 $\genfrac{(}{)}{0pt}{}{m,n}{i,j}$.
$m$ refers to the column number of the unitary cell to which belongs the site
 where the hole-like particle has been created, $n$ refers to the row number of 
 the unitary cell, and $i$ (column), $j$ (row) are indices
 representing the position of that site inside the unitary cell.
$i$ and $j$ run from 0 to 7 and from 0 to 1, respectively.
Now we start to explain with some details the origin of contributions to the
 effective Hamiltonian for a single quasiparticle propagating
 in the spin background depicted by Fig.\ref{fig1} which is
 an exemplification of coexistence between AF and bond orders.
We concentrate on the case of a spin-up quasiparticle.
Within the approximation which we use, operators $h^\dagger_{i \uparrow}$
 create in the vacuum $|\Omega\rangle$ eigenstates
 of the unperturbed Hamiltonian $H_0$,
\begin{equation}
H_0=\sum_{\langle i,j \rangle} S_i S_j 
-t\sum_{\langle i',j' \rangle}c^\dagger_{i \sigma} c_{j \sigma}
+J\sum_{\langle i',j' \rangle} (S^z_i S^z_j -\frac{n_i n_j}{4})
\end{equation}
where $\langle i,j \rangle$ are pairs of sites on which singlets depicted in 
 Fig.\ref{fig1} have been formed and $\langle i',j' \rangle$ are pairs
 of NN sites belonging to AF domains.
Action of $H_0$ is restricted to the space containing states in which none
 of the sites is doubly occupied.
This Hamiltonian contains the Ising part of the exchange energy of links inside 
 domains.
It also drives hole hopping between NN sites inside each domain.
Since all matrix elements which may give rise to deconfinement of a hole,
 have been by definition removed from $H_0$,
 polaron states $|\Psi_i\rangle$ are its eigenstates.
Processes which bring about deconfinement of holes will be treated at the latter
 stage of the calculation as a perturbation.
The exchange energy and the hopping  between sites belonging to bonds occupied
 by singlets in Fig.\ref{fig1} are the only contribution to the Hamiltonian
 $H_0$ from DWs.
Terms in the Hamiltonian of the $tJ$M which couple a site belonging to a 
 ladder-like DW with a site belonging to a domain do not contribute to $H_0$.
It is clear that the vacuum state $|\Omega\rangle$ and the single particle
 states $h^\dagger_{i \sigma}|\Omega\rangle$ are eigenstates of $H_0$.
Within the first order approximation the on-site energy of a quasiparticle with
 spin up created at the site $\genfrac{(}{)}{0pt}{}{m,n}{0,1}$
 which belongs to a  DW is $2J$.
From now on, the reference value of the energy is the energy of the
 vacuum state $|\Omega\rangle$.
The contribution from a destroyed singlet to the on-site energy of a hole-like
 quasiparticle created at the site
 $\genfrac{(}{)}{0pt}{}{m,n}{0,1}$ is $(3/4)J$.
A hole-like quasiparticle occupying the site 
 $\genfrac{(}{)}{0pt}{}{m,n}{0,1}$ with spin 
 up is by definition the same as a single fermion with spin up 
 which occupies the site $\genfrac{(}{)}{0pt}{}{m,n}{0,0}$
 belonging to the bond 
 $\genfrac{(}{)}{0pt}{}{m,n}{0,0}$-$\genfrac{(}{)}{0pt}{}{m,n}{0,1}$.
Since spins on the sites $\genfrac{(}{)}{0pt}{}{m,n}{0,0}$
 and $\genfrac{(}{)}{0pt}{}{m-1,n}{7,0}$ are in this case parallel,
 the lowest order contribution to the exchange energy of the
 bond between these two sites additionally increases by $J/4$ compared
 to the contribution from this bond in the  vacuum state $|\Omega\rangle$. 
In the presence of a hole, the contribution from the potential term 
 $-\sum_{\langle i,j \rangle} n_i n_j/4$ in the initial Hamiltonian
 of the $tJ$M is higher by $J$.
By adding all partial contribution we get the value $2J$ of the total on-site
 energy of the spin-up quasiparticle at the site 
$\genfrac{(}{)}{0pt}{}{m,n}{0,1}$.
The same on-site energy have spin up quasiparticles created in DWs at all sites
 which are NN of sites in domains occupied by spins pointing down.
Finally we deduce that in the effective single-particle Hamiltonian
 appears a term
\begin{eqnarray}
\delta_1 H_{eff}&=&2J \sum_{m,n}
[h^\dagger_{\genfrac{(}{)}{0pt}{}{m,n}{0,1} \uparrow}
 h_{\genfrac{(}{)}{0pt}{}{m,n}{0,1} \uparrow}
+h^\dagger_{\genfrac{(}{)}{0pt}{}{m,n}{1,1} \uparrow}
 h_{\genfrac{(}{)}{0pt}{}{m,n}{1,1} \uparrow}
\nonumber \\
&+&
 h^\dagger_{\genfrac{(}{)}{0pt}{}{m,n}{4,0} \uparrow}
 h_{\genfrac{(}{)}{0pt}{}{m,n}{4,0} \uparrow}
+h^\dagger_{\genfrac{(}{)}{0pt}{}{m,n}{5,0} \uparrow}
 h_{\genfrac{(}{)}{0pt}{}{m,n}{5,0} \uparrow}].
\label{d1heff}
\end{eqnarray}
The fermionic operators $h^\dagger_{\genfrac{(}{)}{0pt}{}{m,n}{i,j}}$ and
 $h_{\genfrac{(}{)}{0pt}{}{m,n}{i,j}}$
 transform the underlying vacuum $|\Omega\rangle$ into the single particle
 state described above and the single particle state into $|\Omega\rangle$,
 respectively.
The notation for indices which we use seems not to be very short, but such
 a form of it is basically unavoidable because the elementary cell of the
 underlying spin background, Fig.\ref{fig1}, is rather big.
That notation also helps to trace easily at the map which is Fig.\ref{fig1},
 the results of hopping events mediated by the Hamiltonian $H_{eff}$.
An analogous term will appear in the Hamiltonian representing a propagating
 spin-down quasiparticle, but with a different set of indices labelling sites
 in the elementary cell. 
The creation of the spin down quasiparticle at these sites, $(0,0)$, $(1,0)$,
 $(4,1)$, $(5,1)$ induces the formation at the ladder leg of an
 uncompensated spin, the direction of which is parallel to the direction
 of the nearest spin in the domain.
For the sake of simplicity we will concentrate in this paper on the propagation
 of the spin-up quasiparticle.
Such a simplification is possible because despite the breakdown of the time 
 reversal symmetry by the underlying spin structure, the energy  of the
 spin-up and the spin-down quasiparticles is degenerate.
Later we will user this observation in the calculation.
The on-site energy of the hole-like quasiparticle is lower for a group of sites
 by $J/2$ then
 for sites to which the contribution (\ref{d1heff}) refers because spin
 of the bond-fermion which appears after a hole has been created at a bond 
 initially occupied by a singlet may be antiparallel to the nearest spin in 
 one of domains,
\begin{eqnarray}
\delta_2 H_{eff}&=&\frac{3}{2} J \sum_{m,n}
[h^\dagger_{\genfrac{(}{)}{0pt}{}{m,n}{0,0} \uparrow}
 h_{\genfrac{(}{)}{0pt}{}{m,n}{0,0} \uparrow}
+h^\dagger_{\genfrac{(}{)}{0pt}{}{m,n}{1,0} \uparrow}
 h_{\genfrac{(}{)}{0pt}{}{m,n}{1,0} \uparrow}
\nonumber \\
&+&
 h^\dagger_{\genfrac{(}{)}{0pt}{}{m,n}{4,1} \uparrow}
 h_{\genfrac{(}{)}{0pt}{}{m,n}{4,1} \uparrow}
+h^\dagger_{\genfrac{(}{)}{0pt}{}{m,n}{5,1} \uparrow}
 h_{\genfrac{(}{)}{0pt}{}{m,n}{5,1} \uparrow}].
\label{d2heff}
\end{eqnarray}
Let us concentrate now on the on-site energy of quasiparticles created at sites
 belonging to domains. 
A spin-up hole-like spin polaron can by definition be created exclusively
 at sites which have been initially occupied by a spin-down fermion.
An obvious contribution to the on-site energy of a hole-like quasiparticle
 energy in domains is $\varepsilon_1$, the minimum value of the matrix element
 (\ref{pH0p}) obtained under the constraint
 $\langle \Psi_i | \Psi_i \rangle=1$.
During the construction of spin polarons we have considered only hopping
 between NN sites, which is governed by the hopping with the highest integral
 $t$.
We have also neglected some ``high order'' processes related to path
 crossing or to the action of the XY term in the Heisenberg model.
In the analysis which we start now, we will discuss in the framework of the
 first order perturbation theory the contribution of some neglected processes
 to the effective Hamiltonian $H_{eff}$.
States depicted by Figs.4(b) and (c) are string states of length 1 and are
 components of the spin-polaron wave-function $|\Psi_i \rangle$ at the site $i$.
A single vertical or horizontal hop of the hole created in the AF background
 at the site $i$ gives rise to the states depicted by Figs.\ref{fig4}(b) 
 and (c) respectively.
The hopping term to next nearest neighbors (NNN) which we treat as a 
 perturbation couples states represented by Figs.4(b) and (c),
 which brings about a contribution to the matrix element 
 $\langle \Psi_i | H_1 | \Psi_i \rangle$, where $H_1=H-H_0$.
This contribution is
\begin{equation}
\gamma_1=4t' \alpha_1^2.
\label{con1}
\end{equation}
We recognize in (\ref{con1}) a product of amplitudes with which string states
 of length 1 appear in the definition (\ref{polar}) of the spin polaron 
 state.
The factor 4 which appears in (\ref{con1}) is related to the fact that hopping
 which couples states (b) and (c) may occur in both directions and that
 analogous coupling takes place between the state depicted by
 Fig.\ref{fig4}(c) and the state depicted by Fig.\ref{fig4}(d).
The state depicted by Fig.\ref{fig4}(d) has been obtained by a single downward
 hop of a hole created in the AF domain at the site $i$.
The contribution 
\begin{equation}
\gamma_2=2t'' \alpha_1^2
\label{con2}
\end{equation}
 to $\langle \Psi_i |H_1|\Psi_i\rangle$ originates in a similar way with
 coupling between states depicted in Fig.\ref{fig4}(c) and (d) by the hopping
 term to third nearest neighbor (TNN) sites.
The coupling between longer string states which are components of the same
 spin-polaron wave-function may also contribute to the renormalization of the
 spin-polaron on-site energy.
To be more specific, the coupling between states as  depicted by
 Figs.\ref{fig4}(e) and (f) and between their
 reflections in the horizontal line running through the site $i$ gives rise
 to the correction
\begin{equation}
\gamma_3=2t' [2 \alpha_2^2+(z-1) \sum_{\mu =3}(z-2)^{\mu-3}\alpha^2_\mu]
\label{con3}
\end{equation}
 to $\langle \Psi_i |H_1|\Psi_i\rangle$, while the coupling between the state 
 depicted by Fig.\ref{fig4}(g) and its reflection in the horizontal line
 running through the site $i$ together with some similar process in which longer
 strings are involved brings about the correction
\begin{equation}
\gamma_4=2t'' [\alpha_2^2+(z-1)\sum_{\mu =3}(z-2)^{\mu-3} \alpha^2_\mu ].
\label{con4}
\end{equation}
The states represented by Figs.\ref{fig4}(e), (f) and (g) have been obtained by
 means of three different sequences of hole moves.
The hole has been created at the site $i$ and these sequences are
 upwards-upwards, upwards-left and left-upwards respectively.
Longer strings pinned to the site $i$ may be coupled in a very 
 similar way, which also gives rise to a change of the on-site energy.
This change has actually already been incorporated into parameters
 $\gamma_3$ and $\gamma_4$.
Finally, by collecting all terms we may infer that in the effective Hamiltonian
 $H_{eff}$ appears the following term referring to the on-site energy
 of spin-polarons
\begin{eqnarray}
\delta_3 H_{eff}&=&(\varepsilon_1+\gamma_1+\gamma_2+\gamma_3+\gamma_4+J/4) 
\nonumber \\
&\times& \sum_{m,n} 
[h^\dagger_{\genfrac{(}{)}{0pt}{}{m,n}{2,1} \uparrow}
 h_{\genfrac{(}{)}{0pt}{}{m,n}{2,1} \uparrow}
+h^\dagger_{\genfrac{(}{)}{0pt}{}{m,n}{3,0} \uparrow}
 h_{\genfrac{(}{)}{0pt}{}{m,n}{3,0} \uparrow}
\nonumber \\
&+&
 h^\dagger_{\genfrac{(}{)}{0pt}{}{m,n}{6,0} \uparrow}
 h_{\genfrac{(}{)}{0pt}{}{m,n}{6,0} \uparrow}
+h^\dagger_{\genfrac{(}{)}{0pt}{}{m,n}{7,1} \uparrow}
 h_{\genfrac{(}{)}{0pt}{}{m,n}{7,1} \uparrow}].
\label{d3heff}
\end{eqnarray}
The additional term $J/4$ in the prefactor is related to the fact that the
 contribution from the contact interaction -$J n_i n_j /4$ between a site $i$
 in a domain and a site $j$ in a DW is higher by 
 $J/4$, when a hole is created in the domain at the site i.
This interaction was neglected, when we were calculating the eigenenergy
 $\varepsilon_1$ of the spin polaron and needs to be taken into account now. 
(\ref{d3heff}) is the last on-site term in the effective Hamiltonian.

The NN hopping integral for the quasiparticle, the propagation of which we 
 describe in this paper is the same as the bare hopping integral provided
 that the sites between which the quasiparticle moves belong to
 the same initially ordered bond inside a DW \cite{wrob04}.
Thus, the term which describes the quasiparticle hopping between NN sites,
 on which singlets have been formed in the underlying spin background, takes
 the following form
\begin{eqnarray}
\delta_4 H_{eff}&=&
-t \sum_{m,n} \{
[h^\dagger_{\genfrac{(}{)}{0pt}{}{m,n}{0,1} \uparrow}
 h_{\genfrac{(}{)}{0pt}{}{m,n}{0,0} \uparrow}
+h^\dagger_{\genfrac{(}{)}{0pt}{}{m,n}{1,1} \uparrow}
 h_{\genfrac{(}{)}{0pt}{}{m,n}{1,0} \uparrow}
\nonumber \\
&+&
 h^\dagger_{\genfrac{(}{)}{0pt}{}{m,n}{4,1} \uparrow}
 h_{\genfrac{(}{)}{0pt}{}{m,n}{4,0} \uparrow}
+h^\dagger_{\genfrac{(}{)}{0pt}{}{m,n}{5,1} \uparrow}
 h_{\genfrac{(}{)}{0pt}{}{m,n}{5,0} \uparrow}]
+H.c.\}.
\nonumber\\
\label{d4heff}
\end{eqnarray}
The NN hopping integral for a pair of sites, which belong to 
 different singlets in the underlying insulating
 state is $t/2$ instead of -$t$.
This change of sign and this reduction of size may be deduced from the
 form of the first term on the right side of Eq.(\ref{hsex}).
Therefore, we may write the following  expression for the contribution to the
 effective Hamiltonian describing quasiparticle hopping between NN sites
 belonging to different bonds on which singlets has been formed in the underlying
 spin background
\begin{eqnarray}
\delta_5 H_{eff}&=&
\frac{t}{2} \sum_{m,n} \{
[h^\dagger_{\genfrac{(}{)}{0pt}{}{m,n}{1,0} \uparrow}
 h_{\genfrac{(}{)}{0pt}{}{m,n}{0,0} \uparrow}
+h^\dagger_{\genfrac{(}{)}{0pt}{}{m,n-1}{0,1} \uparrow}
 h_{\genfrac{(}{)}{0pt}{}{m,n}{0,0} \uparrow}
\nonumber \\
&+&
 h^\dagger_{\genfrac{(}{)}{0pt}{}{m,n}{1,1} \uparrow}
 h_{\genfrac{(}{)}{0pt}{}{m,n}{0,1} \uparrow}
+h^\dagger_{\genfrac{(}{)}{0pt}{}{m,n-1}{1,1} \uparrow}
 h_{\genfrac{(}{)}{0pt}{}{m,n}{1,0} \uparrow}
\nonumber \\
&+&
 h^\dagger_{\genfrac{(}{)}{0pt}{}{m,n}{5,0} \uparrow}
 h_{\genfrac{(}{)}{0pt}{}{m,n}{4,0} \uparrow}
+h^\dagger_{\genfrac{(}{)}{0pt}{}{m,n-1}{4,1} \uparrow}
 h_{\genfrac{(}{)}{0pt}{}{m,n}{4,0} \uparrow}
\nonumber \\
&+&
 h^\dagger_{\genfrac{(}{)}{0pt}{}{m,n}{5,1} \uparrow}
 h_{\genfrac{(}{)}{0pt}{}{m,n}{4,1} \uparrow}
+h^\dagger_{\genfrac{(}{)}{0pt}{}{m,n-1}{5,1} \uparrow}
 h_{\genfrac{(}{)}{0pt}{}{m,n}{5,0} \uparrow}]
+H.c.\}.
\nonumber\\
\label{d5heff}
\end{eqnarray}
The value of the NN hopping integral, between a site which belongs to a
 domain and a site which belongs a DW, can be inferred from the 
 first part of the right side in Eq.(\ref{defs2}),
\begin{eqnarray}
\delta_6 H_{eff}&=&
-\frac{t}{\sqrt{2}} \alpha_0 \sum_{m,n} \{
[h^\dagger_{\genfrac{(}{)}{0pt}{}{m,n-1}{7,1} \uparrow}
 h_{\genfrac{(}{)}{0pt}{}{m,n}{0,1} \uparrow}
\nonumber\\
&+&
 h^\dagger_{\genfrac{(}{)}{0pt}{}{m,n}{2,1} \uparrow}
 h_{\genfrac{(}{)}{0pt}{}{m,n}{1,1} \uparrow}
+h^\dagger_{\genfrac{(}{)}{0pt}{}{m,n}{3,0} \uparrow}
 h_{\genfrac{(}{)}{0pt}{}{m,n}{4,0} \uparrow}
\nonumber\\
&+&
 h^\dagger_{\genfrac{(}{)}{0pt}{}{m,n}{6,0} \uparrow}
 h_{\genfrac{(}{)}{0pt}{}{m,n}{5,0} \uparrow}]
+H.c.\}.
\label{d6heff}
\end{eqnarray}
Since the time reversal symmetry and the translational symmetry are broken
 inside AF domains, the propagating quasiparticle can not move between
 different sublattices and any term related to NN hopping inside domains is not
 generated in the effective Hamiltonian $H_{eff}$.
NNN quasiparticle hopping inside DWs and hopping between sites which belong
 to a DW and a domain is a first order process that is mediated by the NNN
 hopping term in the bare Hamiltonian.
The explicit forms of related contributions to $H_{eff}$ may be deduced
 from formulas (\ref{hsex}), (\ref{defs1}) and (\ref{defs2}),
\begin{eqnarray}
\delta_7 H_{eff}&=&
\frac{t'}{2} \sum_{m,n} \{
[h^\dagger_{\genfrac{(}{)}{0pt}{}{m,n}{1,1} \uparrow}
 h_{\genfrac{(}{)}{0pt}{}{m,n}{0,0} \uparrow}
\nonumber\\
&+&
 h^\dagger_{\genfrac{(}{)}{0pt}{}{m,n-1}{1,1} \uparrow}
 h_{\genfrac{(}{)}{0pt}{}{m,n}{0,0} \uparrow}
+h^\dagger_{\genfrac{(}{)}{0pt}{}{m,n}{1,0} \uparrow}
 h_{\genfrac{(}{)}{0pt}{}{m,n}{0,1} \uparrow}
\nonumber\\
&+&
 h^\dagger_{\genfrac{(}{)}{0pt}{}{m,n-1}{0,1} \uparrow}
 h_{\genfrac{(}{)}{0pt}{}{m,n}{1,0} \uparrow}
+h^\dagger_{\genfrac{(}{)}{0pt}{}{m,n}{5,1} \uparrow}
 h_{\genfrac{(}{)}{0pt}{}{m,n}{4,0} \uparrow}
\nonumber\\
&+&
 h^\dagger_{\genfrac{(}{)}{0pt}{}{m,n-1}{5,1} \uparrow}
 h_{\genfrac{(}{)}{0pt}{}{m,n}{4,0} \uparrow}
+h^\dagger_{\genfrac{(}{)}{0pt}{}{m,n}{5,0} \uparrow}
 h_{\genfrac{(}{)}{0pt}{}{m,n}{4,1} \uparrow}
\nonumber\\
&+&
 h^\dagger_{\genfrac{(}{)}{0pt}{}{m,n-1}{4,1} \uparrow}
 h_{\genfrac{(}{)}{0pt}{}{m,n}{5,0} \uparrow}]
+H.c.\},
\label{d7heff}
\\
\delta_8 H_{eff}&=&
-\frac{t'}{\sqrt{2}} \alpha_0 \sum_{m,n} \{
[h^\dagger_{\genfrac{(}{)}{0pt}{}{m-1,n}{7,1} \uparrow}
 h_{\genfrac{(}{)}{0pt}{}{m,n}{0,0} \uparrow}
\nonumber\\
&+&
 h^\dagger_{\genfrac{(}{)}{0pt}{}{m-1,n-1}{7,1} \uparrow}
 h_{\genfrac{(}{)}{0pt}{}{m,n}{0,0} \uparrow}
+h^\dagger_{\genfrac{(}{)}{0pt}{}{m,n}{2,1} \uparrow}
 h_{\genfrac{(}{)}{0pt}{}{m,n}{0,0} \uparrow}
\nonumber\\
&+&
 h^\dagger_{\genfrac{(}{)}{0pt}{}{m,n-1}{2,1} \uparrow}
 h_{\genfrac{(}{)}{0pt}{}{m,n}{0,0} \uparrow}
+h^\dagger_{\genfrac{(}{)}{0pt}{}{m,n}{3,0} \uparrow}
 h_{\genfrac{(}{)}{0pt}{}{m,n}{4,1} \uparrow}
\nonumber\\
&+&
 h^\dagger_{\genfrac{(}{)}{0pt}{}{m,n+1}{3,0} \uparrow}
 h_{\genfrac{(}{)}{0pt}{}{m,n}{4,1} \uparrow}
+h^\dagger_{\genfrac{(}{)}{0pt}{}{m,n}{6,0} \uparrow}
 h_{\genfrac{(}{)}{0pt}{}{m,n}{5,1} \uparrow}
\nonumber\\
&+&
 h^\dagger_{\genfrac{(}{)}{0pt}{}{m,n+1}{6,0} \uparrow}
 h_{\genfrac{(}{)}{0pt}{}{m,n}{5,1} \uparrow}]
+H.c.\}.
\label{d8heff}
\end{eqnarray}
The task of finding the formula for the term describing NNN hopping inside AF
 domains is little bit more tedious.
For example the coupling by the XY term in the Heisenberg model between
 string states depicted by Fig.\ref{fig4}(f) and (h) gives rise to 
 hopping terms in $H_{eff}$ which shift a spin polaron from the site $i$ to
 the site $j$ and vice versa.
We analyze now the XY term, because it has been neglected during the first
 stage of the analysis, when quasi-confined spin polaron states have been 
 constructed.
Since the string state depicted by Fig.\ref{fig4}(f) is a component of the
 wave function $|\Psi_i\rangle$
 for the spin polaron created at the site $i$ and the string state depicted
 by Fig.\ref{fig4}(h) is evidently a component
 of the wave function for the spin polaron created at the site $j$, the
 coupling between these components gives rise to the coupling between 
 spin polaron states $|\Psi_i\rangle$ and $|\Psi_j\rangle$.
This brings about a contribution to the matrix element
 $\langle \Psi_j|H|\Psi_i\rangle$ and to the hopping 
 term in the effective Hamiltonian $H_{eff}$.
It is probably useful to remind now that $H_{eff}$
 is an approximation to the initial Hamiltonian $H$ in (\ref{hamorg}).
This approximation is expressed in terms of the operators creating and
 annihilating spin polarons.
Also the coupling between the states depicted by Fig.\ref{fig4}(g) and (h)
 contributes to the hopping term between the sites $i$ and $j$ in the 
 effective Hamiltonian $H_{eff}$.
This coupling is mediated by the XY term.
We have already discussed the action of the XY term which by removing two
 defects in the AF structure transforms string states of length 2, tails of
 which are pinned to the site $i$, into the state representing a hole created
 at the site $j$ in the AF ordered domain.
By a string tail we mean its end opposite the end at which sits a hole.
The XY term may also transform a state representing a hole created in the
 domain at the site $i$ into a string state of length 2 pinned at the tail 
 to the site $j$.
This additional coupling between components of the spin polaron states at
 the sites $i$ and $j$ doubles the value of the hopping integral between these
 sites in the effective Hamiltonian.
Thus, after a little thought we may deduce that the integral for the NNN 
 hopping inside domains is
\begin{equation}
\tau_1=2J \sum_{\mu =2} (z-2)^{\mu -2} \alpha_\mu \alpha_{\mu -2}.
\label{tau1}
\end{equation}
The first term in the sum presented above refers to coupling between strings
 of length 0 and 2. 
We have just outlined its origin  in detail.
Other terms appear in the sum (\ref{tau1}) because longer strings, which
 are created when a hole moves further from the site $j$ in Figs.\ref{fig4}(f),
 (g) and (h), are also coupled by the XY term in the Heisenberg model.
The hop left of the hole depicted by Fig.\ref{fig4}(g) gives rise to the string
 state of length three, Fig.\ref{fig4}(i), which is pinned to the site $j$.
Since we neglected the possibility of path crossing when we were constructing
 the quasiconfined spin-polaron states, some corrections need to be made now.
For example we did not consider before, that by applying the NN hopping term
 to the state depicted by Fig.\ref{fig4}(i) we may create the state depicted 
 by Fig.\ref{fig4}(j).
Since the latter state is a string-like component of the spin-polaron at the
 site $j$ obtained by hopping downward and left of a hole created at that site,
 we deduce that the process described above generates the NNN hopping
 term in the effective Hamiltonian with the amplitude
\begin{equation}
\tau_2=2t \alpha_3 \alpha_2.
\label{tau2}
\end{equation}
The factor 2 originates with the fact that the motion of a hole around
 a plaque in the square lattice may take place clockwise and anti-clockwise.
We also recognize $\alpha_3$ and $\alpha_2$ as amplitudes of strings, which 
 have the length 3 and 2, respectively.
Exactly such strings which are components of the spin-polaron states at the
 sites $i$ and $j$ in Figs.\ref{fig4}(i), (j) are coupled by
 the NN hopping term in the initial Hamiltonian.
The NNN term in the initial Hamiltonian generates coupling between states 
 representing ``bare'' holes created in the AF background of domains.
Since these states are also string components with the length 0 of some 
 spin polaron wave functions, coupling between the latter is also generated.
An example of such coupled string states are Figs.\ref{fig4}(h) and (k).
The contribution to the NNN hopping integral in $H_{eff}$ is
\begin{equation}
\tau_3=t' \alpha_0^2.
\label{tau3}
\end{equation}
The NNN term in the bare Hamiltonian also couples states depicted by
 Figs.\ref{fig4}(f) and (l) which are string-like components with length 2
 of spin-polaron states at the sites $i$ and $j$, respectively.
The coupling amplitude is
\begin{equation}
\tau_4=2t' \alpha_2^2.
\label{tau4}
\end{equation}
The appearance of the factor 2 is related to the fact that states which are
 created when holes move between the sites $i$ and $j$ in opposite directions
 around the plaque, than in the case of the states depicted by
 Figs.\ref{fig4}(f) and (l) are also coupled by the bare NNN hopping.
The state in Fig.\ref{fig4}(l) has been obtained by hopping left and downward of
 a hole which has been initially created at the site $j$.
By collecting all contributions, which we have discussed above, we see that
 the new term in to $H_{eff}$ is, 
\begin{eqnarray}
\delta_9 H_{eff}&=&
(\tau_1+\tau_2+\tau_3+\tau_4)
\nonumber \\ &\times&
\sum_{m,n} \{
[h^\dagger_{\genfrac{(}{)}{0pt}{}{m,n}{3,0} \uparrow}
 h_{\genfrac{(}{)}{0pt}{}{m,n}{2,1} \uparrow}
+h^\dagger_{\genfrac{(}{)}{0pt}{}{m,n-1}{2,1} \uparrow}
 h_{\genfrac{(}{)}{0pt}{}{m,n}{3,0} \uparrow}
\nonumber\\ &+&
 h^\dagger_{\genfrac{(}{)}{0pt}{}{m,n}{7,1} \uparrow}
 h_{\genfrac{(}{)}{0pt}{}{m,n}{6,0} \uparrow}
+h^\dagger_{\genfrac{(}{)}{0pt}{}{m,n-1}{7,1} \uparrow}
 h_{\genfrac{(}{)}{0pt}{}{m,n}{6,0} \uparrow}]
+H.c.\}.
\nonumber\\
\label{d9heff}
\end{eqnarray}
By means of a similar analysis, as for the hopping between NNN sites we may 
 find the TNN hopping term in the effective Hamiltonian.
For the operator representing the quasiparticle hopping inside DWs we get
\begin{eqnarray}
\delta_{10} H_{eff}&=&
\frac{t''}{2}\sum_{m,n} \{
[h^\dagger_{\genfrac{(}{)}{0pt}{}{m,n+1}{0,0} \uparrow}
 h_{\genfrac{(}{)}{0pt}{}{m,n}{0,0} \uparrow}
\nonumber\\ &+&
 h^\dagger_{\genfrac{(}{)}{0pt}{}{m,n+1}{0,1} \uparrow}
 h_{\genfrac{(}{)}{0pt}{}{m,n}{0,1} \uparrow}
+h^\dagger_{\genfrac{(}{)}{0pt}{}{m,n+1}{1,0} \uparrow}
 h_{\genfrac{(}{)}{0pt}{}{m,n}{1,0} \uparrow}
\nonumber\\ &+&
 h^\dagger_{\genfrac{(}{)}{0pt}{}{m,n+1}{1,1} \uparrow}
 h_{\genfrac{(}{)}{0pt}{}{m,n}{1,1} \uparrow}
+h^\dagger_{\genfrac{(}{)}{0pt}{}{m,n+1}{4,0} \uparrow}
 h_{\genfrac{(}{)}{0pt}{}{m,n}{4,0} \uparrow}
\nonumber\\ &+&
 h^\dagger_{\genfrac{(}{)}{0pt}{}{m,n+1}{4,1} \uparrow}
 h_{\genfrac{(}{)}{0pt}{}{m,n}{4,1} \uparrow}
+h^\dagger_{\genfrac{(}{)}{0pt}{}{m,n+1}{5,0} \uparrow}
 h_{\genfrac{(}{)}{0pt}{}{m,n}{5,0} \uparrow}
\nonumber\\ &+&
 h^\dagger_{\genfrac{(}{)}{0pt}{}{m,n+1}{5,1} \uparrow}
 h_{\genfrac{(}{)}{0pt}{}{m,n}{5,1} \uparrow}]
+H.c.\}.
\label{d10heff}
\end{eqnarray}
The TNN hopping of the quasiparticle between domains and DWs is governed by the
 following term
\begin{eqnarray}
\delta_{11} H_{eff}&=&
-\frac{t''}{\sqrt{2}} \alpha_0 \sum_{m,n} \{
[h^\dagger_{\genfrac{(}{)}{0pt}{}{m-1,n}{0,0} \uparrow}
 h_{\genfrac{(}{)}{0pt}{}{m,n}{0,0} \uparrow}
\nonumber\\ &+&
 h^\dagger_{\genfrac{(}{)}{0pt}{}{m,n}{2,1} \uparrow}
 h_{\genfrac{(}{)}{0pt}{}{m,n}{0,1} \uparrow}
+h^\dagger_{\genfrac{(}{)}{0pt}{}{m,n}{3,0} \uparrow}
 h_{\genfrac{(}{)}{0pt}{}{m,n}{1,0} \uparrow}
\nonumber\\ &+&
 h^\dagger_{\genfrac{(}{)}{0pt}{}{m-1,n}{7,1} \uparrow}
 h_{\genfrac{(}{)}{0pt}{}{m,n}{1,1} \uparrow}
+h^\dagger_{\genfrac{(}{)}{0pt}{}{m,n}{6,0} \uparrow}
 h_{\genfrac{(}{)}{0pt}{}{m,n}{4,0} \uparrow}
\nonumber\\ &+&
 h^\dagger_{\genfrac{(}{)}{0pt}{}{m,n}{2,1} \uparrow}
 h_{\genfrac{(}{)}{0pt}{}{m,n}{4,1} \uparrow}
+h^\dagger_{\genfrac{(}{)}{0pt}{}{m,n}{3,0} \uparrow}
 h_{\genfrac{(}{)}{0pt}{}{m,n}{5,0} \uparrow}
\nonumber\\ &+&
 h^\dagger_{\genfrac{(}{)}{0pt}{}{m,n}{7,1} \uparrow}
 h_{\genfrac{(}{)}{0pt}{}{m,n}{5,1} \uparrow}]
+H.c.\}.
\label{d11heff}
\end{eqnarray}
To TNN hopping term between sites in domains contribute: a) the mechanism 
 which is based on the shortening of strings by the action of the XY term,
 b) the coupling of strings with length 0 by the TNN hopping term in the
 initial Hamiltonian and c) the exchange between the head and the tail
 of a straight string with length 2.
The last process is  mediated by the TNN hopping term
 in the initial Hamiltonian (\ref{hamorg}).

The contributions to the hopping integral in these three cases are
\begin{eqnarray}
\tau_5 &=&2J \sum_{\mu =2} (z-2)^{\mu -2} \alpha_\mu \alpha_{\mu -2},
\\
\tau_6 &=& t'' \alpha_0^2,
\\
\tau_7 &=& t'' \alpha_2^2.
\label{tau567}
\end{eqnarray}
Since the origin of these couplings is the same as for the NNN hopping
 term in $H_{eff}$ we do not discuss them in detail.
Thus, the TNN hopping operator, which is the last contribution to $H_{eff}$
 discussed by us at the approximation level, that we have assumed, takes
 the form,
\begin{eqnarray}
\delta_{12} H_{eff}&=&(\tau_5+\tau_6+\tau_7)
\sum_{m,n} \{
[h^\dagger_{\genfrac{(}{)}{0pt}{}{m,n+1}{2,1} \uparrow}
 h_{\genfrac{(}{)}{0pt}{}{m,n}{2,1} \uparrow}
\nonumber\\
&+&
h^\dagger_{\genfrac{(}{)}{0pt}{}{m,n+1}{3,0} \uparrow}
 h_{\genfrac{(}{)}{0pt}{}{m,n}{3,0} \uparrow}
+h^\dagger_{\genfrac{(}{)}{0pt}{}{m,n+1}{6,0} \uparrow}
 h_{\genfrac{(}{)}{0pt}{}{m,n}{6,0} \uparrow}
\nonumber \\
&+&
h^\dagger_{\genfrac{(}{)}{0pt}{}{m,n+1}{7,1} \uparrow}
 h_{\genfrac{(}{)}{0pt}{}{m,n}{7,1} \uparrow}]
+H.c.\}.
\label{d12heff}
\end{eqnarray}
Fig.\ref{pardisp} depicts the electronic structure which we have obtained
 by solving the Hamiltonian $H_{eff}$. 
It represents the energy dispersion E({\bf p}) of all bands along the line
 $(0,0)$-$(\pi,0)$
 -$(\pi,\pi)$-$(0,0)$ and the line obtained by performing the rotation by 
 $\pi/2$ around the point $(0,0)$.
Such a combination of dispersion curves is justified by conditions
 in which experiments are performed, because it seems that stripes in some
 regions of the sample may run vertically, while in other regions they may
 run horizontally.
The reason for this mixing can be, for example, sample twinning.
Nondispersing parts of bands are related to some obstacles for hole propagation
 in the directions perpendicular to stripes.
That kind of motion seems to be blocked, which is clear because $H_{eff}$
 by definition can not mediate quasiparticle motion occurring exclusively in
 the direction perpendicular to stripes.
It seems that the lack of the energy dispersion in some directions accounts for
 straight patches of the spectral weight which appear in the density maps
 obtained by means of ARPES measurements \cite{ino02}. 
The coordination frame applied to draw these maps is the same as we have
 used to obtain
 Fig.\ref{pardisp} (momentum-energy), while the second derivative of ARPES
 spectra plays the role of a density-like parameter
 which marks regions with the high spectral weight \cite{ino00}.
We shall make a more detailed comparison with ARPES spectra, after we
 calculate the spectral weight by means of the approach applied by us.
That approach is based, as we have already mentioned several times,
 on a combination of bond and string formalisms.

\section{Single particle spectral function in the stripe phase with coexisting
 bond and AF orders}
\begin{figure}
 \unitlength1cm
\begin{picture}(6.5,9)
\epsfxsize=5.8cm
\put(0,0){\epsfbox{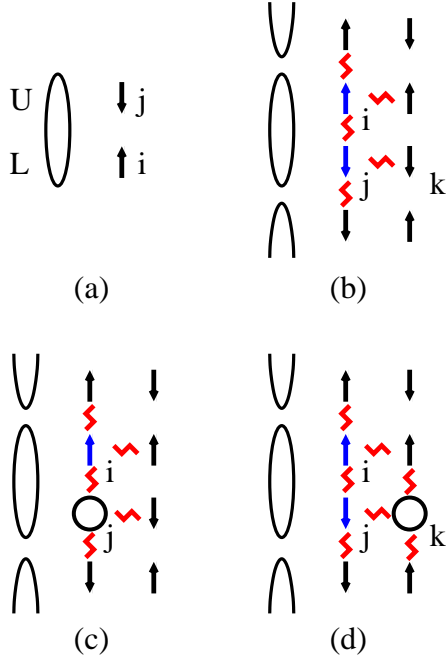}}
\end{picture}
\caption{Graphical representation of states involved in generation
 of quantum fluctuations in $|\Omega\rangle$. Contributions from these
 fluctuations determine to a large extent the shape of the spectral
 function.}
\label{qnfns}
\end{figure}

\begin{figure}
 \unitlength1cm
\begin{picture}(9,5)
\epsfxsize=15cm
\put(0,-13){\epsfbox{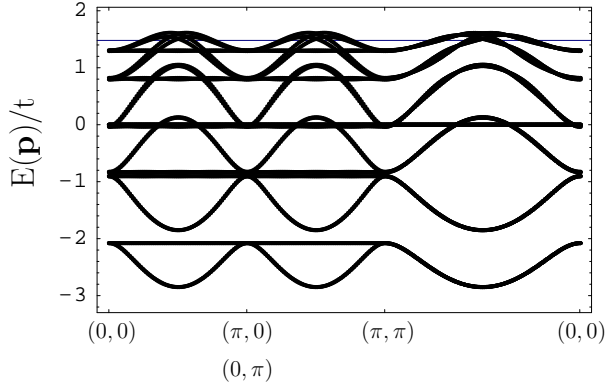}}
\end{picture}
\caption{Band structure in the stripe phase, obtained within the scenario
 of coexisting bond and AF orders.}
\label{pardisp}
\end{figure}

ARPES probes the one-particle spectral function
 $A^- (\mathbf{k},\omega)$.
We neglect in the further analysis of ARPES spectra the influence of
 temperature.
We also assume that the description of the photoelectric effect
 in terms of Fermi's golden rule is sufficient and finally we omit, in
 the calculation, electromagnetic dipole matrix elements between the wave
 function of photoelectron and the wave functions of the electrons in the
 initial states.
Furthermore, we apply the single particle approximation, in the
 calculation of the spectral function  
 $A^- (\mathbf{k},\omega)$, which is natural, because
 the Hamiltonian $H_{eff}$ by definition does not contain the interaction
 terms.
Since electrons are emitted by the photoelectric effect, at $T=0$
 information is gathered only about the one-electron removal part of the
 spectral function, which takes the form
\begin{equation}
A^-(\mathbf{k},\omega)=
\sum_m |\langle\Psi_m^{N-1}|c_{\mathbf{k}}|\Psi^N_i\rangle|^2 
\delta(\omega+E_m^{N-1}-E^N_i).
\label{spfn}
\end{equation}
Due to the formation of the stripe structure the shape of which is
 determined by the underlying
 spin background depicted by Fig.\ref{fig1}, the 1BZ
 gets reduced by a factor of 8 in the horizontal direction and by a factor of
 2 in the vertical direction.
The Hamiltonian $H_{eff}$ may be written in the diagonal form in terms of
 operators $h^\dagger_{\mathbf{k},\uparrow,\alpha}$, and their Hermitian
 conjugates, which are determined by the form of Hamiltonian eigenstates,
\begin{equation}
h^\dagger_{\mathbf{k_R},\uparrow,\alpha}=\frac{1}{\sqrt{NL}}\sum_{n,l}
e^{i\mathbf{k}_R(nd,lw)}
\sum_{i,j} F_{\mathbf{k}_R,(i,j),\uparrow,\alpha}
 h_{\genfrac{(}{)}{0pt}{}{n,l}{i,j} \uparrow}^\dagger
\label{ft}
\end{equation}
 where $\mathbf{k}_R$ belongs to the reduced 1BZ, $\alpha$ labels
 the band number, $d=8$ is the length of the elementary supercell,
 $w=2$ is the width of the elementary supercell and $Nd \times Lw$ 
 is the system size.
Thus, within the single-particle approach, the one-electron removal part of
 the spectral function is approximated as
\begin{eqnarray}
A^-(\mathbf{k},\omega)&=&
\sum_\alpha \delta(\omega+\varepsilon_{\mathbf{k}, \alpha})
\nonumber\\
&\times&
|\sum_{n,l}\sum_{i,j}\sum_{i',j'}F^\ast_{\mathbf{k},mod \mathbf{K}_R,(i,j);
\uparrow,\alpha} 
e^{i\mathbf{k}(nd,lw)}
\nonumber\\
&\times&
e^{-i\mathbf{k}(i',j')}
\langle \Omega | h_{\genfrac{(}{)}{0pt}{}{n,l}{i,j} \uparrow}
c_{\genfrac{(}{)}{0pt}{}{0,0}{i',j'} \downarrow}|\Omega\rangle|^2,
\label{spsfap}
\end{eqnarray}
 where $\mathbf{k}mod \mathbf{K}_R$ denotes the vector $\mathbf{k}$ reduced
 to the 1BZ of the superlattice, the elementary cell of which is depicted by
 Fig.\ref{fig1} and $\varepsilon_{\mathbf{k}, \alpha}$
 is the energy of the $\alpha$-th band in Fig.\ref{pardisp}.
In order to evaluate (\ref{spsfap}) we need to find matrix elements,
\begin{equation}
M_{\genfrac{(}{)}{0pt}{}{n,l}{i,j} \genfrac{(}{)}{0pt}{}{0,0}{i',j'}}=\langle \Omega| 
 h_{\genfrac{(}{)}{0pt}{}{n,l}{i,j} \uparrow}
 c_{\genfrac{(}{)}{0pt}{}{0,0}{i',j'} \downarrow} 
|\Omega\rangle.
\label{matel}
\end{equation}
A scheme showing how to do this in the framework of the spin polaron (string)
 approach was developed before \cite{ederbecker91}.
For example, the removal of the spin down electron at the site $(2,1)$ in the
 elementary cell of the underlying spin background depicted by Fig.\ref{fig1}
 gives rise to a state which is component of the wave function for the polaron
 created  at this site.
Thus the matrix element (\ref{matel}) for $(i',j')=(i,j)=(2,1)$ and
 $(n,l)=0$ is $\alpha_0$.
In the appendix we list all sites labelled by the pairs of numbers 
 $(i',j')$, $(n,l)$, $(i,j)$, for which a nonvanishing matrix element 
 (\ref{matel}) exists.
The removal of a spin down electron from a site which belongs to ladder-like DWs
 in the state depicted by Fig.\ref{fig1} gives rise to the hole-like
 quasiparticle $h$ at this site.
Since this removal may be mediated by the electron annihilation operator
 $c$ in (\ref{matel}), a nonvanishing diagonal matrix element $M=-1/\sqrt{2}$
 is generated.
Quantum spin fluctuation which are generated in the underlying spin background
 schematically depicted by Fig.\ref{fig1} can not be neglected when we evaluate
 the spectral function.
Within the first order perturbation theory the admixture $\delta |\phi\rangle$
 of quantum fluctuations to the groundstate $|\phi_0\rangle$ of $H_0$
 brought about by the perturbation $H_1$ is
\begin{equation}
\delta|\phi\rangle=-\sum_n \frac{\langle \psi_n|H_1|\phi_0\rangle}
{E_n-E_0} |\psi_n\rangle,
\label{1order}
\end{equation}
 where $|\psi_n\rangle$ are excited eigenstates of $H_0$ with energies $E_n$
 and $E_0$ is the groundstate energy. 
The action of the exchange interaction
 between sites depicted by Fig.\ref{qnfns}(a), $L(U)$ belonging to a DW and
 $i(j)$ belonging to an AF domain, may transform the singlet on sites $L$
 and $U$ into a triplet.
Within our approach we treat this part of the exchange interaction as the
 perturbation.
The transformed state is an excited eigenstate of $H_0$ with the energy $J$ and
 also a quantum fluctuation that, according to the formula (\ref{1order}),
 contributes to the underlying vacuum state $|\Omega\rangle$,
 about which we assume that it is an eigenstate of
 the Hamiltonian (\ref{hamorg}).
$|\Omega\rangle$ is schematically depicted
 by Fig.\ref{fig1} in which quantum corrections have not been taken into
 account even within the framework of the first order perturbation theory.
The electron annihilation operator $c$ that acts on a site belonging to a bond,
 which has been excited to the triplet state, transforms it into the state
 representing the hole-like quasiparticle $h$ created at that site.
This process gives rise an addendum to the matrix element (\ref{matel}).
This addendum  may be attributed to the existence of quantum corrections
 to the vacuum state $|\Omega\rangle$.
For example the value of this addendum is -$1/(2\sqrt{2})$ for the matrix
 element (\ref{matel}) labelled by the indices 
$(i',j')=(i,j)=(n,l)=(0,0)$.
The XY part of the exchange interaction which we treat as a perturbation, 
 creates in $|\Omega\rangle$ an excitation that  takes the form of two NN spins
 turned upside down, Fig.\ref{qnfns}(b), with respect to the underlying  spin 
 structure, Fig.\ref{fig1}.
According to the recipe (\ref{1order}) this excitation contributes a correction
 to the underlying vacuum state $|\Omega\rangle$.
If the annihilation operator $c$ removes the spin down fermion at the site $j$
 in Fig.\ref{qnfns}(b), the configuration depicted by Fig.\ref{qnfns}(c)
 will be created, which is a string state, a component of the wave function
 $|\psi_i\rangle$ for the spin polaron at the site $i$.
Since this spin polaron state is by definition created by the fermionic
 operator $h^\dagger$ a contribution -$\alpha_1/4$ to some diagonal matrix
 elements of the form (\ref{matel}) is generated.
We discuss now the last category of processes which give rise to new terms
 in the sum that appears in (\ref{spsfap}).
The electron removal from the site $k$, Fig.\ref{fig4}(b), gives
 rise to a string state of length 2, Fig.\ref{fig4}(d), which is
 a component of the wave function for the spin polaron at the site $i$.
Since the spin-polaron state is created by the operator $h^\dagger$, a 
 nonvanishing matrix element is generated.
Its value is -$\alpha_2/4$, which gets multiplied by a factor of 2, because
 there are two strings of length two, connecting sites $i$ and $k$.
By collecting all contributions to the sum which appears in (\ref{spsfap})
 the one-electron removal part of the spectral function
 $A^-(\mathbf{k}, \omega)$ can be evaluated.
We will apply in the numerical evaluation some Lorentzian broadening
 of the Dirac
 delta function, which is justified because experimental measurements,
 with which we are going to compare our results, have finite resolution
 and also some averaging procedure is often applied to present experimental
 data.

\section{Numerical evaluation of the spectral weight, comparison with results of
 ARPES experiments and concluding remarks}
\begin{figure}
 \unitlength1cm
\begin{picture}(9,4)
\epsfxsize=16cm
\put(-0.5,-13.5){\epsfbox{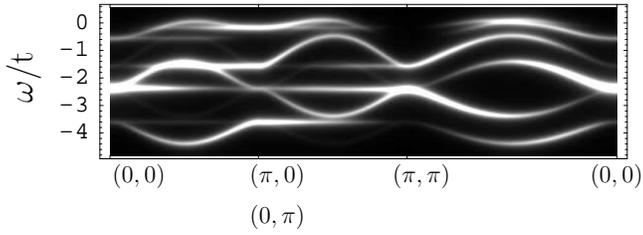}}
\end{picture}
\caption{Spectral weight intensity $A^-(\mathbf{k},\omega)$ along some
 directions at the doping level $15\%$. Lorentzian broadening with
 the width $0.1t$ has been applied.}
\label{parintmap}
\end{figure}

It seems that a most complete set of data to compare with our theoretical
 analysis, provide ARPES measurements of the LSCO 
 and Nd-LSCO systems not for the doping level $x=0.125$ but
 for the doping level $x=0.15$ \cite{ino00,zhou99,zhou01,ino02}.
We will present our results in a way similar to the method of
 presentation, which was used in experimental papers reporting
 these measurements.
The NN hopping parameter $t$ defines a unit in which energy is measured.
We choose $J/t=0.4$, $t'/t=-0.1$, $t''/t=0.05$.
Similar parameters have been applied in a recent theoretical analysis of
 ARPES spectra in the 2D $tJ$M, based on the exact diagonalization of the
 $5 \times 4$ cluster and in a separate calculation performed by means of the
 bond operator formalism applied to the columnar bond order underlying
 spin structure \cite{eder04b}.
This coincidence helps to make comparison between results of the calculation
 presented in our paper and results of the earlier analysis.
The choice of the Hamiltonian parameters is ever to some extent arbitrary.
On the other hand a theoretical analysis of ARPES spectra from LSCO systems,
 based on the tight binding approach indicates that the ratios
 $|t'/t|$ and $|t''/t|$ are lower for these systems than for other members
 of cuprate family \cite{tohyama99}.
Parameters suggested by the authors of the analysis based on a tight binding
 approach \cite{tohyama99} are basically the same as parameters
 applied  in our calculation.
The position of the Fermi energy in the band structure at the doping level
 $15\%$ has been determined in our calculation
 by counting the number of hole-like states.
It has been marked as a narrow straight line in Fig.\ref{pardisp}.
Fig.\ref{parintmap} depicts the intensity of the one-electron removal spectrum
 function presented in the coordination frame momentum-energy.
Contributions from vertical and horizontal stripes have been summed in order
 to account for presumed coexistence of these structures in different parts
 of the sample.
In some agreement with the experimental result obtained at the doping level
 $15\%$ \cite{ino00} we notice in the calculated spectrum
 a strongly dispersing band between points $(0,0)$ and
 $(\pi,0)$/$(0,\pi)$ which approaches the Fermi level, $\omega=0$,
 near the zone boundary,
 where it joins a flat patch formed by the region of high spectral intensity.
After passing the anti-nodal region as a straight narrow strip, the band-like
 region of high intensity disappears somewhere between $(\pi,0)$/$(0,\pi)$ and
 $(\pi,\pi)$ points.
In the calculated spectrum we also notice two cusps near
 $\mathbf{k}=(\pi/2,0)$ and $\mathbf{k}=(\pi,\pi/2)$ which are absent
 in the experimental spectrum.
This discrepancy may be attributed to the fact that to this high-intensity patch
 which looks like a single band in the plot, actually contribute
 two of many bands which may be seen in Fig.\ref{pardisp}.
The band which appears in the results of the exact diagonalization performed
 for the $tJ$M and the results of the calculation based on the scenario of the
 columnar spin-Peierls order \cite{eder04b} has also the strong dispersion but 
 does not flatten in the anti-nodal region.
Despite that neither the spectral weight of the quasiparticle propagating
 in the bond ordered spin background nor the spectral weight of the
 quasiparticle propagating in the AF background \cite{wells}
 resembles ARPES spectra from doped LSCO and doped Nd-LSCO in the stripe
 phase, spectral properties of the model in which these two phases coexist
 gives rise to qualitative agreement with experimental data.
This agreement may be attributed to the fact that the band structure of the
 model of nano-scale phase separation which we discuss here has some
 features which are present in band structures of both pure homogeneous phases.
For example in our results we actually see a remnant of a strip formed by
 high intensity which takes
 the shape of a strongly dispersing band along the line between points $(0,0)$
 and $(\pi,0)$/$(0,\pi)$ and resembles the band that is formed
 in the columnar bond-ordered phase.
This band-like structure actually does not flatten and reaches
 the maximum near the points $(0,\pi /2)$/$(\pi /2,0)$. 
On the other hand it is likely that the presence of this maximum  may be
 attributed to the brute force method of sewing the bond-ordered
 parts of the system with the parts which are AF-ally ordered.
One thing is for sure, that the lack of the straight high intensity patch at
 the anti-nodal region in results of the theory based on the scenario of the 
 columnar spin-Peierls phase demonstrates that the spectral function
 calculated within this picture does not agree in some details with measured
 ARPES spectra and suggests that it is necessary to take into account also
 the long range AF correlations in order to formulate a theory which
 accounts for the spectral properties of the cuprates in the stripe phase.
In the region between the zone center and antinodal points $(\pi,0)$/$(0,\pi)$
 similar agreement has been observed between the
 experimental data and results of a calculation based on the dynamical mean
 field theory (DMFT) applied to the Hubbard model \cite{fleck00}.

Along the line connecting the points $(0,0)$ and $(\pi,\pi)$ both in ARPES
 spectra and in our results we see a band in the vicinity of the Fermi surface.
This band has a maximum at the points $(\pi /2,\pi /2)$, bends downward
 and disappears near the zone corner.
A similar feature may be observed in the results of an exact diagonalization
 performed for the $t$-$t'$-$t''$-$J$ model and in the results of the
 calculation performed within the scenario of the bond-ordered columnar
 phase \cite{eder04b}.
The DMFT of the stripe phase in the Hubbard model at the doping
 level $15\%$ gives rise to a slightly different result, namely
 it seems that the band crosses the Fermi level near the point
 $(\pi /2,\pi /2)$.

\begin{figure}
 \unitlength1cm
\begin{picture}(9,5)
\epsfxsize=4 cm
\put(0,0.2){\epsfbox{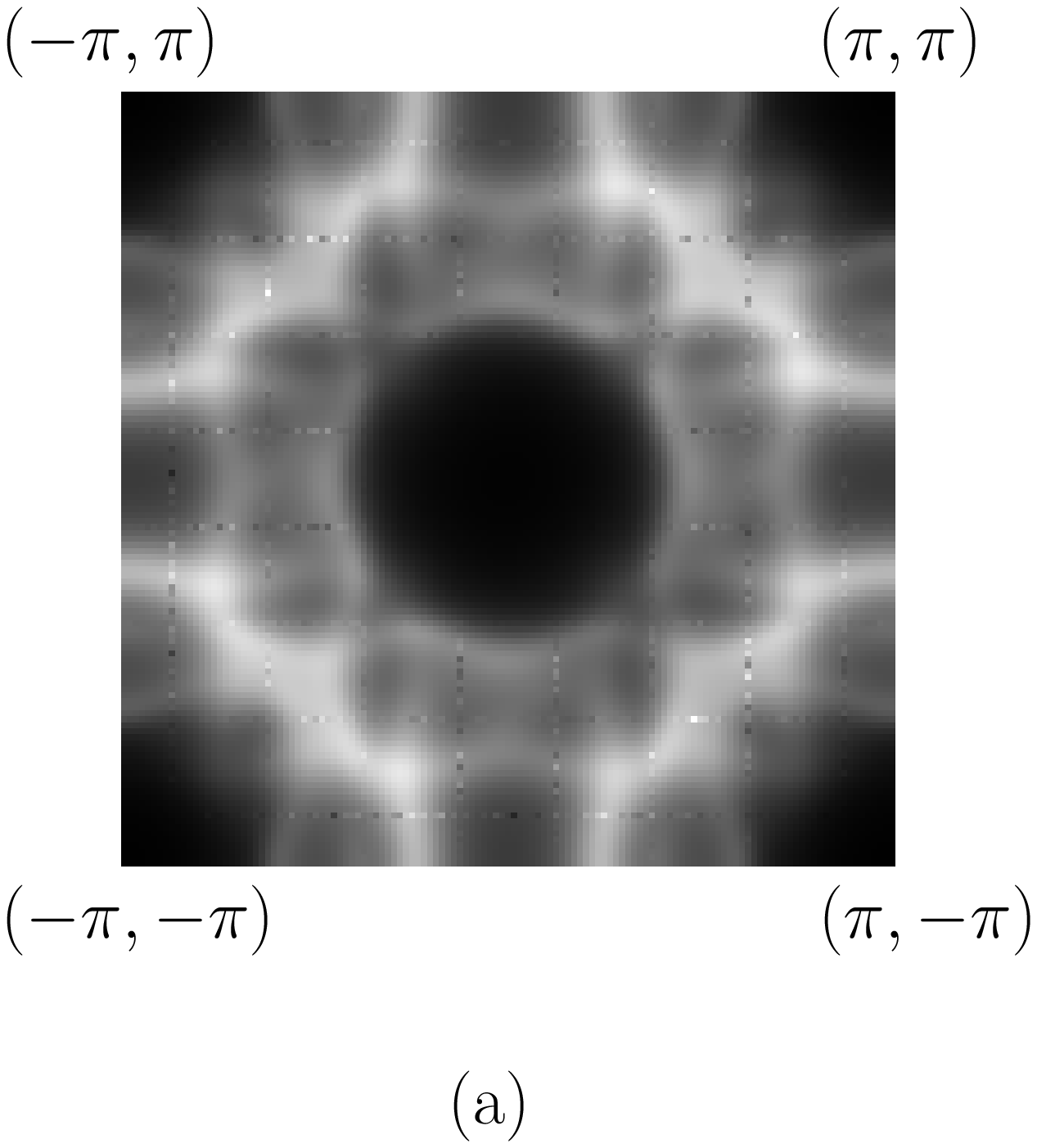}}
\epsfxsize=4 cm
\put(4.5,0.2){\epsfbox{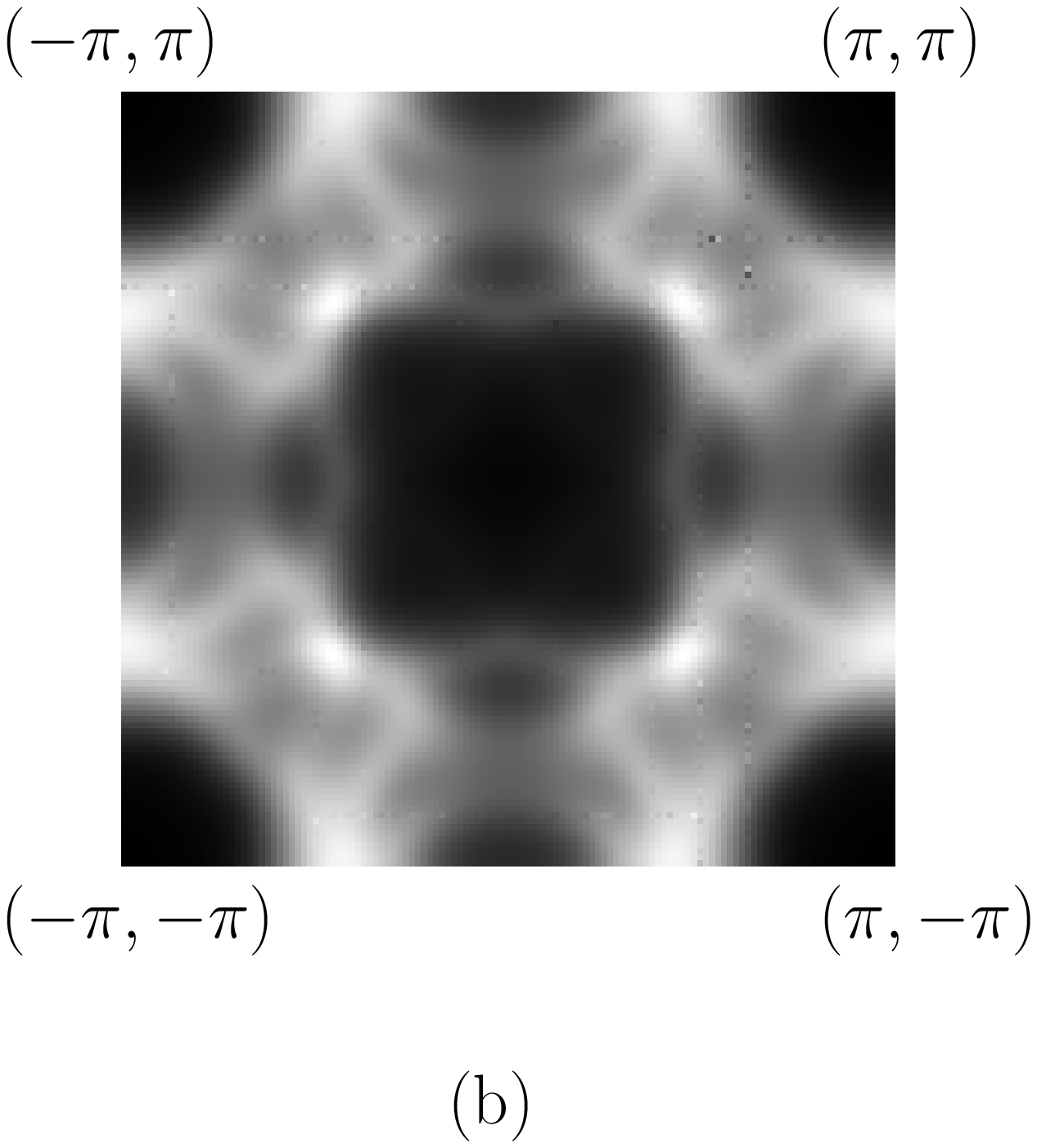}}
\end{picture}
\caption{Intensity map of the spectral weight $A^-(\mathbf{k},\omega)$ at
 $E_F$.}
\label{parflcut}
\end{figure}

The intensity map of the one-electron removal spectral function
 $A^-(\mathbf{k},\omega)$ at the Fermi energy obtained in the framework
 of the scenario which 
 assumes coexistence of bond and AF orders is depicted by Fig.\ref{parflcut}.
Fig.\ref{parflcut}(a) refers to results of the calculation in which we have assumed
 that bond order is formed on legs in ladder-like DWs.
That figure also shows some agreement with ARPES data from LSCO
 and Nd-LSCO \cite{zhou01} obtained for the doping level $15\%$.
We see well developed spectral weight in the nodal regions.
These regions are bridged by high-intensity continuous almost straight
 high-intensity patches.
It seems that the agreement between the experimental results and
 the results of the theory based on the mixture
 of the bond formalism and the spin polaron approach is better
 in this respect than the agreement with
 results of previous calculations based on the CPT applied to microscopic
 models which are the $tJ$M and the Hubbard model \cite{zacher00,zacher02}.
 
Fig.\ref{parflcut}(b) depicts the spectral weight at the Fermi energy
 obtained in a separate calculation for the underlying spin background
 with bond order on rungs.
We know from the results of a previous paper \cite{wrob04} that
 such a structure is less stable and do not expect much similarity
 with experimental results.
Such a lack of similarity may be seen, indeed.
For example the high-intensity patches does not form a shape
 resembling the Fermi surface obtained by means of calculations
 based on the local density approach.
Such a shape may be observed
 both in experimental spectra and in Fig.\ref{parflcut}(a). 
In Fig.\ref{parflcut} apart from patches of high spectral density we notice
 additional regular structures formed by regions of enhanced intensity.
It seems that the origin of thouse structures may be attributed  to simplicity
 of our approach within which fluctuations of the underlying spin structure 
 depicted in Fig.\ref{fig1} are neglected to a great extent.
It is natural to expect that such fluctuations smear out the contribution
 to the spectral function from excitations which may be classified as 
 incoherent background and only the dominating quasiparticle contributions,
 which may be seen as bright patches in Fig.\ref{parflcut}, are preserved in the
 real system.

Calculations based on the phenomenological approach to disordered charge
 stripes and antiphase spin domains give rise to a pattern formed by regions
 of high spectral intensity in the 1BZ which strongly resembles ARPES spectra
 \cite{salkola, granath02}.
Unfortunately no microscopic justification has been provided, of
 phenomenological one-body Hamiltonians which has been applied to derive
 the spectral density by means of calculations based on this scenario.
Our calculation is based on the microscopic $t$-$t'$-$t''$-$J$ model.
On the other hand it seems that disorder may give rise spreading of
 spectral weight over the whole anti-nodal region.
Such a spreading is not observed in the results of our calculation.
In a previous paper we have demonstrated that the magnetic structure
 of the stripe which we have considered here, is likely to have lowest
 energy at and above the doping level $12.5\%$, if the distance between axes
 of nearest stripes is 4 lattice spacings, as it has been suggested by
 experiments \cite{wrob04}.
Bond order parallel to stripe axes and long range AF order coexist in
 this magnetic structure.
It has been suggested that the shape of the Fermi surface seen in ARPES
 spectra from Nd-LSCO and LSCO at the doping level in the range
 $12.5\%-15\% $ may be also explained in the framework of more conventional band
 calculations which neglect the formation of nano-scale
 inhomogeneities \cite{bansilpriv}.
It seems to be hard to reconcile such a way of thinking with the evidence for
 stripes forming in these systems.

In conclusion, motivated by results of a previous calculation \cite{wrob04}
 indicating that, at the doping level $1/8$ and above, the stripe structure,
 which consists of a) hole-filled two-leg ladder-like DWs with the spin-Peierls
 order formed on legs and b) AF domains of width 2 lattice spacings and with the
 changing phase of the sublattice magnetization by $\pi$ across each DW,
 is stable, we have performed the calculation of the single-particle
 spectral density which  is generated in such a system.
Our analysis has been made in the framework of the $t$-$t'$-$t''$-$J$ model,
 with parameter values in the range suggested by comparison between band
 structure calculations and the Fermi surface of overdoped LSCO systems. 
The calculation which we have performed is a combination of the bond
 fermion method and the spin polaron approach.
We observe pronounced spectral weight both in the anti-nodal and nodal
 regions.
Very similar features may be seen in ARPES spectra from LSCO and Nd-LSCO at the 
 filling level $15\%$, which is exactly the same as we have assumed in the
 calculation.
This similarity is not trivial, because different optional structures of bond
 order in DWs, as the spin-Peierls order on rungs in the ladder-like DWs
 or the bond order which takes the shape of two layers in a 
 brick-wall, give rise to spectra at the Fermi level which have completely
 different forms.
We consider the observed agreement between experiment and theory as an 
 argument for the scenario of coexisting bond and long range AF 
 orders in the stripe phase of doped cuprates.

\acknowledgments
PW acknowledges partial support by the Polish Science Committee (KBN) under
 contract No 2P03B00925.

\appendix*
\section{}
We list here contributions to matrix elements (\ref{matel}).
The contribution to (\ref{matel}) is $\alpha_0$ for the following values of
 indices $(i',j')$, $(n,l)$, $(i,j)$:
 (2,1), (0,0), (2,1); (3,0), (0,0), (3,0); 
 (4,0), (0,0), (4,0); (5,1), (0,0), (5,1).
The contribution is -$\frac{1}{\sqrt{2}}$ for
 (0,0), (0,0), (0,0); (0,1), (0,0), (0,1); 
 (1,0), (0,0), (1,0); (1,1), (0,0), (1,1);
 (4,0), (0,0), (4,0); (4,1), (0,0), (4,1);
 (5,0), (0,0), (5,0); (5,1), (0,0), (5,1);
-$\frac{1}{2\sqrt{2}}$ for
 (0,0), (0,0), (0,0); (1,0), (0,0), (1,0);
 (4,1), (0,0), (4,1); (5,1), (0,0), (5,1);
$\frac{1}{2\sqrt{2}}$ for
 (0,1), (0,0), (0,1); (1,1), (0,0), (1,1);
 (4,0), (0,0), (4,0); (5,0), (0,0), (5,0);
-$\frac{\alpha_1}{4}$ for
 (2,0), (0,0), (2,1); (2,0), (0,0), (3,0);
 (2,0), (0,-1), (2,1); (3,1), (0,0), (2,1);
 (3,1), (0,0), (3,0); (3,1), (0,1), (3,0); 
 (6,1), (0,0), (6,0); (6,1), (0,0), (7,1);
 (6,1), (0,1), (6,0); (7,0), (0,0), (6,0);
 (7,0), (0,0), (7,1); (7,0), (0,-1), (7,1);
-$\frac{\alpha_2}{2}$ for
 (2,1), (0,0), (3,0); (2,1), (0,1), (3,0);
 (3,0), (0,0), (2,1); (3,0), (0,-1), (2,1);
 (6,0), (0,0), (7,1); (6,0), (0,-1), (7,1);
 (7,1), (0,0), (6,0); (7,1), (0,1), (6,0);
-$\frac{\alpha_2}{4}$ for
 (2,1), (0,1), (2,1); (2,1), (0,-1), (2,1);  
 (3,0), (0,1), (3,0); (3,0), (0,-1), (3,0);
 (6,0), (0,1), (6,0); (6,0), (0,-1), (6,0);
 (7,1), (0,1), (7,1); (7,1), (0,-1), (7,1).


\newpage

\end{document}